\documentclass[12pt, draftclsnofoot, onecolumn]{IEEEtran}
\usepackage{multicol}
\usepackage{widetext}
\usepackage{amsmath}
\usepackage{amssymb}
\usepackage{graphicx}
\usepackage{float}
\usepackage{fixltx2e}
\usepackage{subeqnarray}
\usepackage{cases}
\allowdisplaybreaks
\IEEEoverridecommandlockouts
\usepackage[square, comma, sort&compress, numbers]{natbib}
\usepackage{multirow}
\usepackage{stfloats}
\usepackage{fixltx2e}
\usepackage{pmat}
\usepackage{balance}
\usepackage{arydshln}
\usepackage{flushend}
\usepackage{subcaption}

\ifCLASSINFOpdf
\else
\fi
\usepackage{amsmath}
\newenvironment{smallarray}[1]
{\null\,\vcenter\bgroup\scriptsize
	\arraycolsep=.13885em
	\hbox\bgroup$\array{@{}#1@{}}}
{\endarray$\egroup\egroup\,\null}

\usepackage{pdfpages}

\newtheorem{theorem}{Theorem}
\newtheorem{lemma}{Lemma}

\begin{document}
%
\title{Asynchronous Channel Training in Multi-Cell Massive MIMO}
%
%
%

\author{\IEEEauthorblockN{Xun Zou, {\it Student Member, IEEE,} and Hamid Jafarkhani, {\it Fellow, IEEE}}%
	\thanks{Results in this paper were presented in part at the IEEE Global Communications Conference (Globecom) 2016 \cite{xun}. This work was supported in part by the NSF Award CCF-1526780.}%
	\thanks{X. Zou and H. Jafarkhani are with the Center for Pervasive Communications and Computing, Department of Electrical Engineering and Computer Science, University of California, Irvine, CA, 92697 USA (email: \{xzou4, hamidj\}@uci.edu).}}

\maketitle

\begin{abstract}
Pilot contamination has been regarded as the main bottleneck in time division duplexing (TDD) multi-cell massive multiple-input multiple-output (MIMO) systems. The pilot contamination problem cannot be addressed with large-scale antenna arrays. We provide a novel asynchronous channel training scheme to obtain precise channel matrices without the cooperation of base stations. The scheme takes advantage of sampling diversity by inducing intentional timing mismatch. Then, the linear minimum mean square error (LMMSE) estimator and the zero-forcing (ZF) estimator are designed. Moreover, we derive the minimum square error (MSE) upper bound of the ZF estimator. In addition, we propose the equally-divided delay scheme which under certain conditions is the optimal solution to minimize the MSE of the ZF estimator employing the identity matrix as pilot matrix. We calculate the uplink achievable rate using maximum ratio combining (MRC) to compare asynchronous and synchronous channel training schemes. Finally, simulation results demonstrate that the asynchronous channel estimation scheme can greatly reduce the harmful effect of pilot contamination. 
\end{abstract}


%
\IEEEpeerreviewmaketitle

\section{Introduction}
%
%
%
%
Massive multiple-input multiple-output (MIMO) technology has been considered as one of the key technologies in next generation wireless communication systems \cite{marzetta2016fundamentals}. Compared with traditional MIMO, base station (BS) is equipped with a large-scale antenna array possessing hundreds or even thousands of antenna elements in massive MIMO systems. The extensive number of antennas provides several benefits. For example, a large-scale antenna array at base station (BS) provides larger antenna gain and higher spectral and energy efficiency \cite{marzetta2016fundamentals}. In addition, a large number of antennas results in orthogonal channel matrices among different users, which is called ``favorable propagation." As a result of the favorable propagation, the small-scale fading in channel gradually disappears, i.e., ``channel hardening" occurs, as the number of antennas grows to infinity \cite{hochwald2004multiple}.\par

However, pilot contamination restricts the performance of multi-cell massive MIMO systems \cite{marzetta2016fundamentals}. Pilot contamination is caused by pilot reuse among different cells. In uncooperative multi-cell systems with an unlimited number of antennas, the effects of uncorrelated noise and fast fading vanish while the inter-cellular interference caused by pilot contamination still exists. It acts as the major bottleneck for overall capacity.\par

Generally, there are two approaches to tackle this issue. One popular approach is based on cooperation among different BSs or clusters. For example, pilot contamination precoding (PCP) has been used in \cite{jose2011pilot} to reduce pilot contamination. Its main idea is to linearly combine messages aimed at terminals from different cells using the same pilot sequence. Another coordinated approach using second-order statistical information about user channels is presented in \cite{yin2013coordinated}. It is demonstrated that the pilot contamination effect vanishes completely under certain conditions on the channel covariance. Moreover, pilot scheduling methods are studied in \cite{jin2015pilot} and \cite{chen2016low}. However, cooperation or coordination among BSs demands high-rate message exchanges and suffers from the propagation delay in backhaul links. The other approach to mitigate the effect of pilot contamination is based on statistical techniques or signal processing methods, such as random pilot design \cite{kapetanovic2013detection}, blind pilot decontamination \cite{muller2014blind}, and group-blind detection \cite{ferrante2016group}, but such statistical methods have high complexity and are valid only when a huge number of antennas is used.\par

 
In this paper, we propose a novel asynchronous channel training scheme based on an oversampling method to acquire accurate channel matrices. We intentionally create timing mismatch between different received signals. While asynchronous channel training method in \cite{kobayashi2015cooperative} has resulted in erroneous estimated channel because of noise and channel correlation, we show how to avoid the drawbacks by designing an appropriate sampling method. It has been shown that the performance of MIMO systems with intentional timing offset between transmitters can be improved by using sampling diversity \cite{das2011mimo,ganji,poorkasmaei2015asynchronous,avendi2015differential}. In this work, we design the linear minimum mean square error (LMMSE) and zero-forcing (ZF) estimators using multiple samples. Furthermore, we propose an equally-divided delay scheme based on the ZF estimator, in which all user delays are equally divided within the symbol length. We claim that, under certain conditions, it is the optimal case for the mean square error (MSE) performance of the ZF estimator. Moreover, we derived a universal expression of the MSE upper bound of the ZF estimator. In addition, we compute the uplink achievable rate employing maximum ratio combining (MRC) as criterion to examine the performance of different channel estimation methods. Finally, our simulation results verify that our scheme is able to mitigate the effect of pilot contamination without BS cooperation or coordination, and the equally-divided delay scheme can provide similar performance as the globally optimal scheme.\par

The main results in our work are based on the assumption that the BS knows the time delay of each received signal. We make such an ideal assumption to study the effects of time delays on the performance. The conventional wisdom in this field suggests that one should try to synchronize all received signals. We show that there is a benefit in designing asynchronous signals, motivated by the increased capacity \cite{verdu1989capacity}. To reach such a counter-intuitive conclusion, we need to assume all time delays are known to find an upper bound on how much gain, if any, an asynchronous signal design can provide. Obviously, the next step, as a future work, is how to estimate the delays or how to deal with unknown delays. In fact, controlling the delays is not an impossible task. The delay of each received signal is the sum of two components: (i) the time delay introduced at the transmitter and (ii) the time delay of the channel. The first component can be controlled by adjusting the transmitter clock while the second component can be measured and sent back to the transmitter through the control channel.  Also, it has been shown, for different set-ups, that the sampling diversity gains are achievable even if the time delays are not known \cite{avendi2015differential}. \par

The rest of this paper is organized as follows. In Section II, we describe the asynchronous channel training system model. In Section III, we introduce LMMSE and ZF estimator design based on the asynchronous channel training scheme. In Section IV, we derive expressions for the MSE upper bound of the ZF estimator. Furthermore, we present the equally-divided delay scheme which is the optimal delay scheme for the ZF estimator in Section V. The calculation of uplink achievable rate is given in Section VI. We present numerical results in Section VII. Finally, we draw our conclusions in Section VIII.\par

Notations: We use bold font variables to denote matrices and vectors. $(\cdot)^H$ denotes the Hermitian transpose, $(\cdot)^T$ denotes the transpose, $tr(\cdot)$ denotes the trace operation, $(\cdot)^{-1}$ denotes the inverse operation, and $\left \|\cdot \right\|_F$ denotes the Frobenius-norm. $\mathbb{E}[\cdot]$ stands for expectation. $\lceil a\rceil$ denotes the smallest integer bigger than $a$. Finally, $x\ \mathrm{mod}\ y$ denotes the modulus after dividing $x$ by $y$. $\mathbf{I}$ represents the identity matrix and $\mathbf{J}$ is the symmetric elementary matrix where all secondary diagonal entries are ones and other elements are zeros. $\ast$ stands for linear convolution operation. The sign function is defined as $\mathrm{sign}(x)=1$ if $x>0$ and $\mathrm{sign}(x)=-1$ if $x<0$

\section{System Model}

\begin{figure}[t b]
	\centering
	\includegraphics[width=3in]{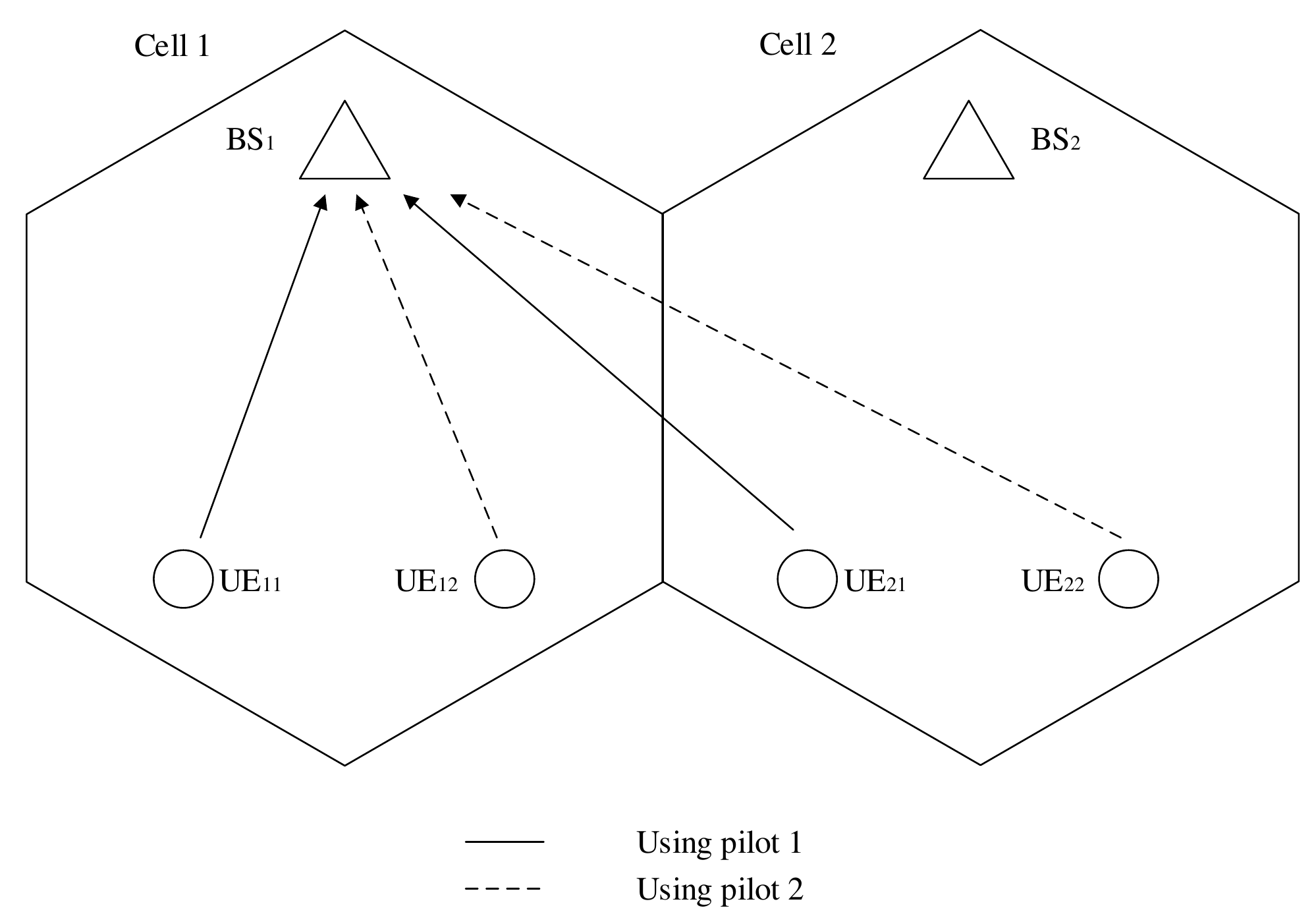}
	\caption{The simplest 2-cell 2-user massive MIMO scenario.}
	\label{scenario}
\end{figure}

In this section, we will present system models for a single-carrier channel training system. In the following discussion, we consider time division duplexing (TDD) massive MIMO systems. Each BS is equipped with $M$ antennas ($M\rightarrow\infty$) while each UE has a single antenna. 

\subsection{2-cell 2-user Scenario}
First, we consider the simplest 2-cell 2-user scenario. There are two cells with two user equipments (UEs) in each cell, as is shown in Fig. \ref{scenario}. Two orthogonal pilots $\mathbf{P}_1$ and $\mathbf{P}_2$ ($1\times L$ vectors) are reused in each cell where $\mathbf{P}_1 \mathbf{P}_2^T=0$. The length of each pilot sequence is equal to the number of pilot sequences $L=2$. UE$_{kn}$ denotes the $n$th UE in the $k$th cell using the $n$th pilot sequence $\mathbf{P}_n$. For UE$_{kn}$, employing time-limited pulse shapes $s_{kn}(t)$, the transmitted signal can be written as
\begin{equation}
\begin{aligned}
X_{kn}(t)=\sum_{i=0}^{L-1}p_n[i]s_{kn}(t-iT),
\end{aligned}
\end{equation} 
in which $k,n\in \{1,2\}$ and $p_n[i]$ is the $i$th symbol of pilot sequence $\mathbf{P}_n$.\par

\begin{figure}[t b]
	\centering
	\includegraphics[width=3.5in]{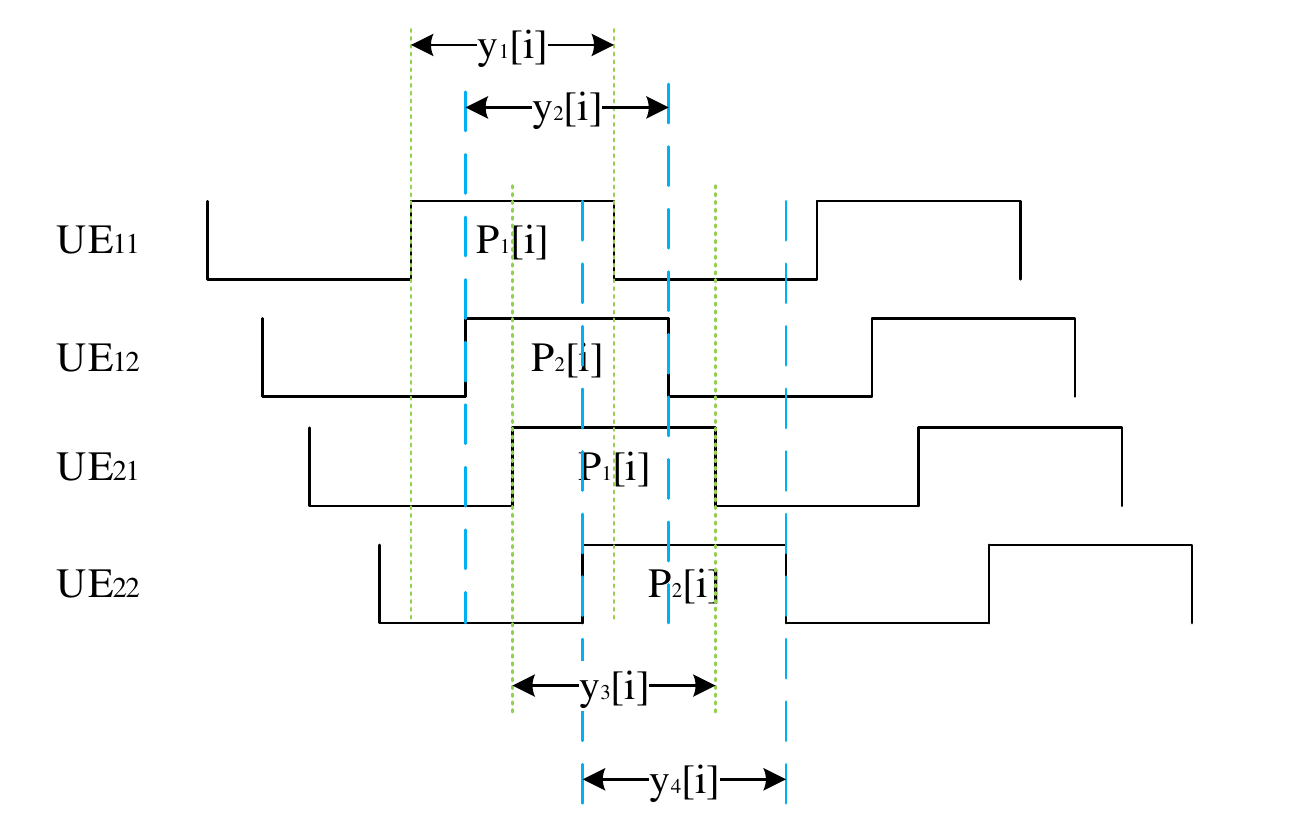}
	\caption{The oversampling procedure for 4 asynchronous received signals.}
	\label{sampling}
\end{figure}

In uplink training, each UE transmits its pilot sequence to the BS to estimate its channel state. The channel vector of UE$_{kn}$, denoted by $\mathbf{h}_{kn}$, is an $M\times 1$ vector whose entries are assumed to be independent and identically distributed (i.i.d.) zero-mean complex Gaussian random variables $\mathbf{h}_{kn}\sim\mathcal{CN}(0,\sigma_{kn}^2\mathbf{I}_M)$. We also assume that the channel vector remains constant for a duration of $L$ pilot symbols. Since each signal $X_{kn}$ goes through a different path, we assume that the signal from UE$_{kn}$ has its own random time delay $\tau_{kn}$, i.e., $\tau_{k_1n_1} \neq \tau_{k_2n_2}$ if $k_1\neq k_2$ or $n_1 \neq n_2$. The relative time delay between UE$_{k_1n_1}$ and UE$_{k_2n_2}$ is defined as $\tau_{k_2n_2}^{k_1n_1}=\tau_{k_1n_1}-\tau_{k_2n_2},\ k_1, k_2, n_1, n_2\in \{1,2\}$. Without loss of generality, it is assumed that $0=\tau_{11}^{11}<\tau_{11}^{12}<\tau_{11}^{21}<\tau_{11}^{22}<T$ where $T$ is the symbol duration. For simplicity of illustration, the transmitted signals are depicted with a rectangular pulse shape in Fig. \ref{sampling} though our approach works for any pulse shape. As explained before, we assume BSs know the time delay of each received signal to find out if an asynchronous signal design is beneficial. The matched filter output is sampled multiple times, each time synchronized by one of the transmitters, i.e., at every $iT+\tau_{kn}$, $k,n\in {1,2}$ and $i=0,1,2,\dots, L-1$. As a result, the number of samples is equal to the number of individual received signals although the BS receives one signal which is the sum of all transmitted signals. \par

Note that the above oversampling does not increase the sampling rate since the sampling rate is $1/T$ for each $y_l[i]$ signal in Fig. \ref{sampling}. Moreover, only one matched filter is needed to realize oversampling at each receiver antenna. Compared with the traditional sampling method, signals are sampled at multiple moments employing oversampling. Accordingly, oversampling does not increase hardware complexity or limit resources available per UE. \par


The first sample at BS$_1$ at time $i\ (1\leq i\leq L-1)$ is given by
\begin{equation}
\begin{aligned}
\mathbf{y}_{11}[i]=&\sqrt\gamma[\mathbf{h}_{11}\ \mathbf{h}_{12}\ \mathbf{h}_{21}\ \mathbf{h}_{22}]\begin{bmatrix}
\rho_{11}^{11}p_1[i] \\ \rho_{11}^{12}p_2[i-1]+\rho_{12}^{11}p_2[i] \\  \rho_{11}^{21}p_1[i-1]+\rho_{21}^{11}p_1[i] \\ \rho_{11}^{22}p_2[i-1]+\rho_{22}^{11}p_2[i]
\end{bmatrix} +\mathbf{n}_{11}[i],
\end{aligned}
\label{eq1}
\end{equation} 
where $\gamma$ is the received signal-to-noise ratio (SNR), $\mathbf{n}_{11}[i]\sim\mathcal{CN}(\mathbf{0},\mathbf{I}_M)$ is the $M\times 1$ additive white Gaussian noise (AWGN) vector, $\rho$ represents the coefficient between two asynchronous signals:
\begin{equation}
\begin{aligned}
\rho^{k_1n_1}_{k_2n_2}&=\int s_{k_1n_1}(t)s_{k_2n_2}(t-\tau_{k_2n_2}^{k_1n_1})dt =  1 - |\tau_{k_2n_2}^{k_1n_1}|/T,\\
\rho^{k_2n_2}_{k_1n_1}&=\int s_{k_1n_1}(t)s_{k_2n_2}(t+\mathrm{sign}(\tau_{k_2n_2}^{k_1n_1})T-\tau_{k_2n_2}^{k_1n_1})dt = |\tau_{k_2n_2}^{k_1n_1}|/T,
\end{aligned}
\end{equation}
where $\mathrm{sign}(x)=1$ if $x>0$ and $\mathrm{sign}(x)=-1$ if $x<0$. At time $i=0$, its sample is defined as
\begin{equation}
\begin{aligned}
\mathbf{y}_{11}[0]=\sqrt\gamma[\mathbf{h}_{11}\ \mathbf{h}_{12}\ \mathbf{h}_{21}\ \mathbf{h}_{22}]\begin{bmatrix}
\rho_{11}^{11}p_1[0] \\ \rho_{12}^{11}p_2[0] \\  \rho_{21}^{11}p_1[0] \\ \rho_{22}^{11}p_2[0]
\end{bmatrix}+\mathbf{n}_{11}[0].
\end{aligned}
\end{equation}
\par
Then, the first sample vector from time 0 to $L-1$ can be written as
\begin{equation}
\begin{aligned}
&\mathbf{Y}_{11}=[\mathbf{y}_{11}[0]\ \mathbf{y}_{11}[1]\ \dots\ \mathbf{y}_{11}[L-1]]\\
&=\sqrt\gamma[\mathbf{h}_{11}\ \mathbf{h}_{12}\ \mathbf{h}_{21}\ \mathbf{h}_{22}]\begin{bmatrix}
\rho_{11}^{11}\mathbf{P}_1 \\ \rho_{11}^{12}\mathbf{P}_2[-1]+\rho_{12}^{11}\mathbf{P}_2 \\  \rho_{11}^{21}\mathbf{P}_1[-1]+\rho_{21}^{11}\mathbf{P}_1 \\ \rho_{11}^{22}\mathbf{P}_2[-1]+\rho_{22}^{11}\mathbf{P}_2
\end{bmatrix}+\mathbf{N}_{11},
\end{aligned}
\label{sample1}
\end{equation}
where $\mathbf{N}_{11}=[\mathbf{n}_{11}[0]\ \mathbf{n}_{11}[1]\ \dots\ \mathbf{n}_{11}[L-1]]$ and $\mathbf{P}_i[k]\ (k\in\{-1, 1\})$ is the pilot sequence $\mathbf{P}_i$ shifted by $k$. For instance, $\mathbf{P}_1[-1]=[0\ p_1[0]\ p_1[1]\ \dots \ p_1[L-2]]$.\par

As for the second, third and fourth samples at BS$_1$, similarly, we have
\begin{equation}
\begin{aligned}
\mathbf{Y}_{12}=\sqrt\gamma[\mathbf{h}_{11}\ \mathbf{h}_{12}\ \mathbf{h}_{21}\ \mathbf{h}_{22}]\begin{bmatrix}
\rho_{12}^{11}\mathbf{P}_1+\rho_{11}^{12}\mathbf{P}_1[1] \\ \rho_{12}^{12}\mathbf{P}_2 \\  \rho_{12}^{21}\mathbf{P}_1[-1]+\rho_{21}^{12}\mathbf{P}_1 \\ \rho_{12}^{22}\mathbf{P}_2[-1]+\rho_{22}^{12}\mathbf{P}_2
\end{bmatrix}+\mathbf{N}_{12}
\end{aligned}
\label{sample2}
\end{equation}
\begin{equation}
\begin{aligned}
\mathbf{Y}_{21}=\sqrt\gamma[\mathbf{h}_{11}\ \mathbf{h}_{12}\ \mathbf{h}_{21}\ \mathbf{h}_{22}]\begin{bmatrix}
\rho_{21}^{11}\mathbf{P}_1+\rho_{11}^{21}\mathbf{P}_1[1] \\ \rho_{21}^{12}\mathbf{P}_2+\rho_{12}^{21}\mathbf{P}_2[1] \\ \rho_{21}^{21}\mathbf{P}_1 \\   \rho_{21}^{22}\mathbf{P}_2[-1]+\rho_{22}^{21}\mathbf{P}_2
\end{bmatrix}+\mathbf{N}_{21}
\end{aligned}
\label{sample3}
\end{equation}
\begin{equation}
\begin{aligned}
\mathbf{Y}_{22}=\sqrt\gamma[\mathbf{h}_{11}\ \mathbf{h}_{12}\ \mathbf{h}_{21}\ \mathbf{h}_{22}]\begin{bmatrix}
\rho_{22}^{11}\mathbf{P}_1+\rho_{11}^{22}\mathbf{P}_1[1] \\ \rho_{22}^{12}\mathbf{P}_2+\rho_{12}^{22}\mathbf{P}_2[1] \\  \rho_{22}^{21}\mathbf{P}_1+\rho_{21}^{22}\mathbf{P}_1[1] \\
\rho_{22}^{22}\mathbf{P}_2
\end{bmatrix}+\mathbf{N}_{22}.
\end{aligned}
\label{sample4}
\end{equation}
\indent Combining Eqs. (\ref{sample1}), (\ref{sample2}), (\ref{sample3}), and (\ref{sample4}) together, we derive 
\begin{equation}
	\begin{aligned}
	\mathbf{Y}&=[\mathbf{Y}_{11}\ \mathbf{Y}_{12}\ \mathbf{Y}_{21}\ \mathbf{Y}_{22}]=\sqrt\gamma\underbrace{[\mathbf{h}_{11}\ \mathbf{h}_{12}\ \mathbf{h}_{21}\ \mathbf{h}_{22}]}_{\mathbf{H}}\cdot\\
	&\underbrace{\begin{bmatrix}
		\rho_{11}^{11}\mathbf{P}_1\ &\rho_{12}^{11}\mathbf{P}_1+\rho_{11}^{12}\mathbf{P}_1[1]\ &\rho_{21}^{11}\mathbf{P}_1+\rho_{11}^{21}\mathbf{P}_1[1]\ &\rho_{22}^{11}\mathbf{P}_1+\rho_{11}^{22}\mathbf{P}_1[1]\\ \rho_{11}^{12}\mathbf{P}_2[-1]+\rho_{12}^{11}\mathbf{P}_2\ &\rho_{12}^{12}\mathbf{P}_2\ &\rho_{21}^{12}\mathbf{P}_2+\rho_{12}^{21}\mathbf{P}_2[1]\ &\rho_{22}^{12}\mathbf{P}_2+\rho_{12}^{22}\mathbf{P}_2[1]\\  \rho_{11}^{21}\mathbf{P}_1[-1]+\rho_{21}^{11}\mathbf{P}_1\ &\rho_{12}^{21}\mathbf{P}_1[-1]+\rho_{21}^{12}\mathbf{P}_1\ &\rho_{21}^{21}\mathbf{P}_1\ &\rho_{22}^{21}\mathbf{P}_1+\rho_{21}^{22}\mathbf{P}_1[1]\\ \rho_{11}^{22}\mathbf{P}_2[-1]+\rho_{22}^{11}\mathbf{P}_2\ &\rho_{12}^{22}\mathbf{P}_2[-1]+\rho_{22}^{12}\mathbf{P}_2\ &\rho_{21}^{22}\mathbf{P}_2[-1]+\rho_{22}^{21}\mathbf{P}_2\ &\rho_{22}^{22}\mathbf{P}_2
		\end{bmatrix}}_{\mathbf{R}}\\
	&+\underbrace{[\mathbf{N}_{11}\ \mathbf{N}_{12}\ \mathbf{N}_{21}\ \mathbf{N}_{22}]}_{\mathbf{N}}.
	\end{aligned}
	\label{wideeq}
	\end{equation}
	\begin{equation}
	\begin{aligned}
	\mathbf{R}=\underbrace{\begin{bmatrix}
		\mathbf{P}_1\ &\ &\ &\ \\
		\  &\mathbf{P}_2\ &\ &\ \\
		\  &\ &\mathbf{P}_1\ &\ \\
		\  &\  &\ &\mathbf{P}_2\  
		\end{bmatrix}}_\mathbf{P}\cdot
	\underbrace{\begin{pmat}[{.|.|.|.}]
		\rho_{11}^{11}\ &0\ &\rho_{12}^{11}\ &0\ &\rho_{21}^{11}\ &0\ &\rho_{22}^{11}\ &0\cr 0\ &\rho_{11}^{11}\ &\rho_{11}^{12}\ &\rho_{12}^{11}\ &\rho_{11}^{21}\ &\rho_{21}^{11}\ &\rho_{11}^{22}\ &\rho_{22}^{11}\cr\-
		\rho_{12}^{11}\  &\rho_{11}^{12}\ &\rho_{12}^{12}\ &0\ &\rho_{21}^{12}\ &0\ &\rho^{12}_{22}\ &0\cr
		0\ &\rho_{12}^{11}\ &0\ &\rho_{12}^{12}\ &\rho_{12}^{21}\ &\rho^{12}_{21}\ &\rho_{12}^{22}\ &\rho^{12}_{22}\cr\-
		\rho_{21}^{11}\ &\rho^{21}_{11}\ &\rho^{12}_{21}\ &\rho_{12}^{21}\  &\rho_{21}^{21}\ &0\ &\rho_{22}^{21}\ &0\cr 
		0\ &\rho^{11}_{21}\ &0\ &\rho_{21}^{12}\  &0\ &\rho_{21}^{21}\ &\rho^{22}_{21}\ &\rho_{22}^{21}\cr\- 
		\rho_{22}^{11}\ &\rho_{11}^{22}\ &\rho_{22}^{12}\ &\rho_{12}^{22}\  &\rho_{22}^{21}\  &\rho_{21}^{22}\ &\rho_{22}^{22}\ &0\cr
		0\ &\rho_{22}^{11}\ &0\ &\rho_{22}^{12}\  &0\ &\rho_{22}^{21}\ &0\ &\rho_{22}^{22}\cr
		\end{pmat}}_{\mathbf{R_P}}.
	\end{aligned}
	\end{equation} 
\indent We rewrite Eq. (\ref{wideeq}) more compactly in the following matrix format:
\begin{equation}
\begin{aligned}
\mathbf{Y}&=\sqrt\gamma\mathbf{H}\mathbf{R}+\mathbf{N} = \sqrt{\gamma}\mathbf{HPR_P + N},
\end{aligned}
\label{simfinal}
\end{equation}
where $\mathbf{N}$ is an additive zero-mean, unit-variance Gaussian noise with covariance matrix:
\begin{equation}
\begin{aligned}
\mathbf{R_{NN}}=\mathbb{E}\left[\frac{\mathbf{N}^H\mathbf{N}}{M}\right]=\begin{bmatrix}
\mathbf{I}_L\ &\mathbf{R}_{12}^{11}\ &\mathbf{R}_{21}^{11}\ &\mathbf{R}_{22}^{11}\\
\mathbf{R}_{11}^{12}\ &\mathbf{I}_{L}\ &\mathbf{R}_{21}^{12}\ &\mathbf{R}_{22}^{12}\\
\mathbf{R}_{11}^{21}\ &\mathbf{R}_{12}^{21}\ &\mathbf{I}_{L}\ &\mathbf{R}_{22}^{21}\\
\mathbf{R}_{11}^{22}\ &\mathbf{R}_{12}^{22}\ &\mathbf{R}_{21}^{22}\ &\mathbf{I}_{L}\\
\end{bmatrix},
\end{aligned}
\end{equation}
where $M$ is the number of antennas in BS, $\mathbf{I}_L$ is the $L\times L$ identity matrix and $\mathbf{R}_{k_2n_2}^{k_1n_1}$ is the covariance matrix of noise matrices $\mathbf{N}_{k_1n_1}$ and $\mathbf{N}_{k_2n_2}$. The noise $\mathbf{N}$ in Eq. (\ref{simfinal}) is correlated because the over-sampled $\mathbf{Y}$ signals have overlaps. For example,
\begin{equation}
\begin{aligned}
\mathbf{R}^{11}_{12}&=\mathbb{E}\left[\frac{\mathbf{N}_{11}^H\mathbf{N}_{12}}{M}\right]=\begin{bmatrix}
\rho^{11}_{12} &0\\
\rho^{12}_{11} &\rho^{11}_{12}
\end{bmatrix}\\
\mathbf{R}_{11}^{12}&=\mathbb{E}\left[\frac{\mathbf{N}_{12}^H\mathbf{N}_{11}}{M}\right]=\begin{bmatrix}
\rho^{11}_{12} &\rho^{12}_{11}\\
0 &\rho^{11}_{12}
\end{bmatrix}.
\end{aligned}
\end{equation}

Note that, for the unit-power pulse shape, i.e., $\int s^2(t)dt=1$, 
\begin{equation}
\begin{aligned}
\mathbf{R_{NN}}=\mathbf{R_P}.
\end{aligned}
\label{RNN=RP}
\end{equation}

If signals from UE$_{11}$ and UE$_{21}$ are synchronous, samples $\mathbf{Y}_{11}$ and $\mathbf{Y}_{21}$ are identical. We have three different samples, i.e., $\mathbf{Y} = [\mathbf{Y}_{11}\ \mathbf{Y}_{12}\ \mathbf{Y}_{22}]$ and $\mathbf{R}$ degrades to
\begin{equation}
\begin{aligned}
\mathbf{R} = \begin{bmatrix}
\rho_{11}^{11}\mathbf{P}_1\ &\rho_{12}^{11}\mathbf{P}_1+\rho_{11}^{12}\mathbf{P}_1[1]\  &\rho_{22}^{11}\mathbf{P}_1+\rho_{11}^{22}\mathbf{P}_1[1]\\ 
\rho_{11}^{12}\mathbf{P}_2[-1]+\rho_{12}^{11}\mathbf{P}_2\ &\rho_{12}^{12}\mathbf{P}_2\  &\rho_{22}^{12}\mathbf{P}_2+\rho_{12}^{22}\mathbf{P}_2[1]\\
\rho_{11}^{11}\mathbf{P}_1\ &\rho_{12}^{11}\mathbf{P}_1+\rho_{11}^{12}\mathbf{P}_1[1]\  &\rho_{22}^{11}\mathbf{P}_1+\rho_{11}^{22}\mathbf{P}_1[1]\\ 
\rho_{11}^{22}\mathbf{P}_2[-1]+\rho_{22}^{11}\mathbf{P}_2\ &\rho_{12}^{22}\mathbf{P}_2[-1]+\rho_{22}^{12}\mathbf{P}_2\  &\rho_{22}^{22}\mathbf{P}_2
\end{bmatrix},
\end{aligned}
\label{R_syn_hybrid}
\end{equation}
where the pilot sequences from UE$_{11}$ and UE$_{21}$ are overlapping and $\mathbf{R}$ is not full-rank, the first and the third rows of $\mathbf{R}$ are identical. This leads to the ``pilot contamination" phenomenon, i.e., the channel vector of target UE is corrupted by UEs reusing the same pilot sequence in other cells. \par

If all four signals are synchronous, there will be only one sample in total. The matrix $\mathbf{R}$ will degrade to an $NK\times L$ matrix, i.e., 
\begin{equation}
\begin{aligned}
\mathbf{R}=\begin{bmatrix}
\mathbf{P}_1\\ \mathbf{P}_2\\ \mathbf{P}_1\ \\ \mathbf{P}_2
\end{bmatrix}.
\end{aligned}
\label{R_syn}
\end{equation}\par

In such a case, $\mathbf{RP}^T_1=[1\ 0\ 1\ 0]^T$, and multiplying $\mathbf{Y}_{M\times L}$ by $\mathbf{P}_1^T$ will result in removing $\mathbf{h}_{12}$ and $\mathbf{h}_{22}$ from the equations while $\mathbf{h}_{11}$ and $\mathbf{h}_{21}$ remain but cannot be separated. Similarly, $\mathbf{h}_{12}$ and $\mathbf{h}_{22}$ cannot be separated if we multiply $\mathbf{Y}$ by $\mathbf{P}_2^T$. \par

Eqs. (\ref{R_syn_hybrid}) and (\ref{R_syn}) demonstrate that our model can also be applied to partially-synchronous and totally-synchronous systems resulting in pilot contamination. 

\subsection{Multi-cell Multi-user Scenario}
In this part, we extend the above model to much more complicated multi-cell multi-user scenarios. We assume that there are $K$ cells each serving $N$ UEs. Then, each BS will have $NK$ samples at each time. There are $N$ orthogonal pilot sequences in total which is the same as the number of UEs per cell, i.e., $\mathbf{P}_n,\ n=1,2,\dots,N$. The time delay of signal from UE$_{kn}$ must not exceed $T$, i.e., $\tau_{kn}\in (0,T)$.\par
\indent In this scenario, Eq. (\ref{simfinal}) still holds, although the matrix dimensions will change as follows:
\begin{equation}
\begin{aligned}
\mathbf{Y}&=[\mathbf{Y}_{11}\ \dots\ \mathbf{Y}_{1K}\ \dots\ \mathbf{Y}_{N1}\ \dots \mathbf{Y}_{NK}]_{M\times NKL}\\
\mathbf{H}&=[\mathbf{h}_{11}\ \dots\ \mathbf{h}_{1K}\ \dots\ \mathbf{h}_{N1}\ \dots \mathbf{h}_{NK}]_{M\times NK}\\
\mathbf{N}&=[\mathbf{N}_{11}\ \dots\ \mathbf{N}_{1K}\ \dots\ \mathbf{N}_{N1}\ \dots \mathbf{N}_{NK}]_{M\times NKL}.
\end{aligned}
\end{equation}
\indent Matrix $\mathbf{R}$ in the multi-cell multi-user scenario is more complicated and will be derived next. The $(r,t)$th block is the signal component from UE$_{k_2n_2}$ at the output of the matched filter synchronized with the signal from UE$_{k_1n_1}$, where $k_1=\lceil r/N\rceil$, $k_2=\lceil t/N\rceil$, $n_1=r\ \mathrm{mod}\ N+1$, and $n_2=t\ \mathrm{mod}\ N+1$. Specifically, when $r=t$, the $(r,t)$th block of $\mathbf{R}$ is given by $\rho_{k_1,n_1}^{k_1,n_1}\mathbf{P}_{n_1}$. If $r\neq t$, we have 
\begin{equation}
\begin{aligned}
\mathbf{R}[r,t]=\rho^{k_1n_1}_{k_2n_2}\mathbf{P}_{n_2}+\rho_{k_1n_1}^{k_2n_2}\mathbf{P}_{n_2}[\mathrm{sign}(\tau_{k_2n_2}^{k_1n_1})].
\end{aligned}
\end{equation}
\indent We also need to derive the covariance matrix of noise, i.e., $\mathbf{R_{NN}}$. For $r=t$, we can write the $(r,t)$th block of $\mathbf{R_{NN}}$ as $\mathbf{I}_L$. For $r\neq t$, the block $\mathbf{R_{NN}}[r,t]$ is the noise covariance matrix of $\mathbf{N}_{k_1n_1}$ and $\mathbf{N}_{k_2n_2}$, i.e., $\mathbf{R}_{k_2n_2}^{k_1n_1}=\mathbb{E}\left\{\frac{\mathbf{N}_{k_1n_1}^H\mathbf{N}_{k_2n_2}}{M}\right\}$. Although $\mathbf{R_{NN}}[r,t]$ is similar to an $L\times L$ circulant matrix, it is not circulant. If $\tau_{k_1n_1}<\tau_{k_2n_2}$,
\begin{equation}
\begin{aligned}
\mathbf{R}_{k_2n_2}^{k_1n_1}&=\mathbb{E}\left[\frac{\mathbf{N}_{k_1n_1}^H\mathbf{N}_{k_2n_2}}{M}\right] =\left[\begin{smallmatrix}
\rho_{k_2n_2}^{k_1n_1}\ &0\ &0\ &\dots\ &0\ &0\ &0\\
\rho^{k_2n_2}_{k_1n_1}\ &\rho_{k_2n_2}^{k_1n_1}\ &0\ &\dots\ &0\ &0\ &0\\
\vdots\ &\vdots\ &\vdots\ &\vdots\ &\vdots\ &\vdots\ &\vdots\\
0\ &0\ &0\ &\dots\ &\rho^{k_2n_2}_{k_1n_1}\ &\rho_{k_2n_2}^{k_1n_1}\ &0\\ 
0\ &0\ &0\ &\dots\ &0\ &\rho^{k_2n_2}_{k_1n_1}\ &\rho_{k_2n_2}^{k_1n_1} 
\end{smallmatrix}\right].
\end{aligned}
\label{lowertriangle}
\end{equation}
\indent If $\tau_{k_1n_1}>\tau_{k_2n_2}$, 
\begin{equation}
\begin{aligned}
&\mathbf{R}_{k_2n_2}^{k_1n_1}=\mathbb{E}\left[\frac{\mathbf{N}_{k_1n_1}^H\mathbf{N}_{k_2n_2}}{M}\right] =\left[\begin{smallmatrix}
\rho_{k_2n_2}^{k_1n_1}\ &\rho^{k_2n_2}_{k_1n_1}\ &0\ &0\ &\dots\ &0\ &0\\
0\ &\rho_{k_2n_2}^{k_1n_1}\ &\rho^{k_2n_2}_{k_1n_1}\ &0\ &\dots\ &0\ &0\\
0\ &0\ &\rho_{k_2n_2}^{k_1n_1}\ &\rho^{k_2n_2}_{k_1n_1}\ &\dots\ &0\ &0\\
\vdots\ &\vdots\ &\vdots\ &\vdots\ &\vdots\ &\vdots\ &\vdots\\
0\ &0\ &0\ &0\ &\dots\ &\rho_{k_2n_2}^{k_1n_1}\ &\rho^{k_2n_2}_{k_1n_1}\\
0\ &0\ &0\ &0\ &\dots\ &0\ &\rho_{k_2n_2}^{k_1n_1} 
\end{smallmatrix}\right].
\end{aligned}
\label{uppertriangle}
\end{equation}
\par

%

\section{Estimator Design}
\subsection{LMMSE Estimator}
In this section, we introduce the LMMSE estimator for single-carrier systems. Note that, according to Eq. (\ref{RNN=RP}), the covariance matrix of noise, $\mathbf{R_{NN}}$, is equal to the coefficient matrix $\mathbf{R_P}$. In this subsection, we use $\mathbf{R_P}$ to substitute $\mathbf{R_{NN}}$. Assume that the LMMSE estimator is $\mathbf{X}_{\mathrm{LMMSE}}$, which can be expressed as 
\begin{equation}
\begin{aligned}
\mathbf{X}_{\mathrm{LMMSE}} &=\sqrt\gamma \mathbf{R}_{\mathbf{P}}^{-1}\mathbf{R}^H(\gamma\mathbf{R}\mathbf{R}_\mathbf{P}^{-1}\mathbf{R}^H+\mathbf{R}_{\mathbf{HH}}^{-1})^{-1}=\sqrt{\gamma} \mathbf{P}^H(\gamma \mathbf{PR_PP}^H + \mathbf{R_{HH}^{-1}})^{-1}.
\end{aligned}
\label{mse}
\end{equation}
\indent In massive MIMO systems, the covariance matrix of $\mathbf{H}$ is diagonal under certain conditions\cite{ngo2014aspects}: 
\begin{equation}
\begin{aligned}
\mathbf{R_{HH}}=\mathbb{E}\left\{\frac{\mathbf{H}^H\mathbf{H}}{M}\right\}=\begin{bmatrix}
\begin{smallmatrix}
\sigma_{11}^2&&&&&&\\
&\ddots &&&&&\\
&&\sigma_{1K}^2&&&&\\
&&&\ddots&&&\\
&&&&\sigma_{N1}^2&&\\
&&&&&\ddots&\\
&&&&&&\sigma_{NK}^2
\end{smallmatrix}
\end{bmatrix},
\end{aligned} 
\end{equation}
where $\sigma_{nk}^2$ is the variance of entries in $\mathbf{h}_{nk}$.

\indent The MSE can be calculated as

\begin{equation}
\begin{aligned}
\mathrm{MSE_{LMMSE}}=&\mathbb{E}\left\{\frac{|| \mathbf{\hat{H}}_{\mathrm{LMMSE}}-\mathbf{H} ||_F^2}{KNM}\right\}
=\frac{1}{KN\gamma}tr\left[\left(\mathbf{PR_{P}P}^H + \frac{1}{\gamma}\mathbf{R_{HH}}^{-1} \right)^{-1}\right].
\end{aligned}
\label{LMMSEestimator}
\end{equation}
\indent Pilot contamination happens in the synchronous system because the received signals from the users sharing the same pilot sequence are mixed. However, by inducing time mismatch and the proposed sampling method, we have more equations for each pilot. This leads to sampling diversity which provides extra gain for channel estimation. The asynchronous channel training scheme takes advantage of such sampling diversity to achieve better performance compared to the synchronous scheme.
\par

\subsubsection{ZF Estimator}
Derived from Eq. (\ref{LMMSEestimator}), the ZF estimator can be written as
\begin{equation}
\begin{aligned}
\mathrm{\mathbf{X}}_{\mathrm{ZF}} = \sqrt{\gamma}\mathrm{\mathbf{P}}^{H}(\gamma\mathrm{\mathbf{PR_PP}}^H)^{-1}.
\end{aligned}
\end{equation}
\indent The MSE using the ZF estimator can be calculated as
\begin{equation}
\begin{aligned}
\mathrm{MSE}_{\mathrm{ZF}} &= \mathbb{E}\left\{\frac{||\mathrm{\mathbf{\hat{H}}}_{\mathrm{ZF}} - \mathrm{\mathbf{H}}||^2_F}{MKN}\right\}
= \frac{1}{KN\gamma}\ tr[(\mathrm{\mathbf{PR_PP}}^H)^{-1}].
\end{aligned}
\label{zfmse}
\end{equation}

\par

\section{MSE Upper Bound for ZF Estimator}
In this section, we analyze the performance of the ZF estimator in single-carrier systems. Let us denote $\mathrm{\mathbf{PR_PP}}^H$ in Eq. (\ref{zfmse}) $\mathrm{\mathbf{A}}$ and write it as
\begin{equation}
\begin{aligned}
&\mathrm{\mathbf{A}} = \left[\begin{smallmatrix}
\mathrm{\mathbf{P}}_1\mathrm{\mathbf{R}}^{11}_{11}\mathrm{\mathbf{P}}_1^H\ &\dots\ &\mathrm{\mathbf{P}}_1\mathrm{\mathbf{R}}^{11}_{1N}\mathrm{\mathbf{P}}_N^H\ &\dots\ &\mathrm{\mathbf{P}}_1\mathrm{\mathbf{R}}^{11}_{K1}\mathrm{\mathbf{P}}_1^H\ &\dots\ &\mathrm{\mathbf{P}}_1\mathrm{\mathbf{R}}^{11}_{KN}\mathrm{\mathbf{P}}_N^H\\
\vdots\ &\vdots\ &\vdots\ &\vdots\ &\vdots\ &\vdots\ &\vdots\\
\mathrm{\mathbf{P}}_N\mathrm{\mathbf{R}}^{1N}_{11}\mathrm{\mathbf{P}}_1^H\ &\dots\ &\mathrm{\mathbf{P}}_N\mathrm{\mathbf{R}}^{1N}_{1N}\mathrm{\mathbf{P}}_N^H\ &\dots\ &\mathrm{\mathbf{P}}_N\mathrm{\mathbf{R}}^{1N}_{K1}\mathrm{\mathbf{P}}_1^H\ &\dots\ &\mathrm{\mathbf{P}}_N\mathrm{\mathbf{R}}^{1N}_{KN}\mathrm{\mathbf{P}}_N^H\\
\vdots\ &\vdots\ &\vdots\ &\vdots\ &\vdots\ &\vdots\ &\vdots\\
\mathrm{\mathbf{P}}_1\mathrm{\mathbf{R}}^{K1}_{11}\mathrm{\mathbf{P}}_1^H\ &\dots\ &\mathrm{\mathbf{P}}_1\mathrm{\mathbf{R}}^{K1}_{1N}\mathrm{\mathbf{P}}_N^H\ &\dots\ &\mathrm{\mathbf{P}}_1\mathrm{\mathbf{R}}^{K1}_{K1}\mathrm{\mathbf{P}}_1^H\ &\dots\ &\mathrm{\mathbf{P}}_1\mathrm{\mathbf{R}}^{K1}_{KN}\mathrm{\mathbf{P}}_N^H\\
\vdots\ &\vdots\ &\vdots\ &\vdots\ &\vdots\ &\vdots\ &\vdots\\
\mathrm{\mathbf{P}}_N\mathrm{\mathbf{R}}^{KN}_{11}\mathrm{\mathbf{P}}_1^H\ &\dots\ &\mathrm{\mathbf{P}}_N\mathrm{\mathbf{R}}^{KN}_{1N}\mathrm{\mathbf{P}}_N^H\ &\dots\ &\mathrm{\mathbf{P}}_N\mathrm{\mathbf{R}}^{KN}_{K1}\mathrm{\mathbf{P}}_1^H\ &\dots\ &\mathrm{\mathbf{P}}_N\mathrm{\mathbf{R}}^{KN}_{KN}\mathrm{\mathbf{P}}_N^H
\end{smallmatrix}\right],
\end{aligned}
\label{A}
\end{equation}
where the ($i$, $j$)th entry denotes the coefficient of signals from UE$_{k_1n_1}$ and UE$_{k_2n_2}$ in which $k_1 = \lfloor\frac{i}{N}\rfloor + 1$, $k_2 = \lfloor\frac{j}{N}\rfloor + 1$, $n_1 = (i - 1) \mathrm{mod} N + 1$ and $n_2 = (j - 1) \mathrm{mod} N + 1$. The ($i$, $j$)th entry can be written as $\mathrm{\mathbf{P}}_{n_1}\mathrm{\mathbf{R}}^{k_1n_1}_{k_2n_2}\mathrm{\mathbf{P}}^H_{n_2}$. Recall that $\mathrm{\mathbf{R}}^{k_1n_1}_{k_2n_2}$ is an identity matrix if $k_1 = k_2$ and $n_1 = n_2$. Therefore, all diagonal entries are 1.\par

Next, we prove the following lemma:
\begin{lemma}
	$\mathrm{\mathbf{A}}$ is a positive definite matrix.
\end{lemma}
\begin{IEEEproof}
	For any non-zero $1\times KN$ vector $\mathrm{\mathbf{x}}$,
	\begin{equation}
	\begin{aligned}
	\mathrm{\mathbf{xAx}}^H &= \mathrm{\mathbf{xPR_PP}}^H\mathrm{\mathbf{x}}^H = \frac{1}{M}\mathbb{E}\{\mathrm{\mathbf{xPN}}^H\mathrm{\mathbf{N}}\mathrm{\mathbf{P}}^H\mathrm{\mathbf{x}}^H\}\\
	& = \frac{1}{M}\mathbb{E}\{||\mathrm{\mathbf{N}}\mathrm{\mathbf{P}}^H\mathrm{\mathbf{x}}^H||^2\} > 0,\ \mathrm{if}\ \mathbf{x} \neq \mathbf{0}.
	\end{aligned}
	\end{equation}
	\indent Therefore, $\mathrm{\mathbf{A}}$ is a positive definite matrix.
\end{IEEEproof}

It is simple to obtain $tr(\mathbf{A}) = \sum_{i} 1/\lambda_i \leqslant dim(\mathbf{A})/\lambda_{min}(\mathbf{A})$ where $\lambda_{min}(\mathbf{A})$ is the minimum eigenvalue of $\mathbf{A}$ and $dim(\mathbf{A})$ is the dimension of $\mathbf{A}$. As a result, the MSE upper bound for the ZF estimator is 
\begin{equation}
	\begin{aligned}
	\mathrm{MSE}_{\mathrm{ZF}} < \frac{1}{\gamma\lambda_{\mathrm{min}}(\mathrm{\mathbf{A}})}.
\end{aligned}
\end{equation}

\section{Equally-divided Delays}
In this section, we present an equally-divided delay scheme based on the ZF estimator. We employ the identity matrix as the pilot matrix since the performance using the identity pilot matrix is better than that of the system with DFT pilot matrix, another commonly used pilot matrix. The identity pilot matrix is also simple to implement. In the following, $\mathbf{P}_n$ represents the $1 \times N$ vector whose $n$th element is 1 and the others are 0.

Revisiting Eq. (\ref{A}), each element of $\mathbf{A}$ can be expressed as $\mathrm{\mathbf{P}}_{n_1}\mathrm{\mathbf{R}}^{k_1n_1}_{k_2n_2}\mathrm{\mathbf{P}}^H_{n_2}$. The elements where $n_1 \neq n_2$ stand for the interference among users using distinct pilot sequences while the elements where $n_1 = n_2$ represent interference from pilot contamination. Due to time asynchrony, the perfect orthogonality among distinct pilot sequences may be damaged. The following theorem claims that the orthogonality will be maintained under certain conditions.


\begin{theorem}
	We separate users in groups according to the pilot sequence they use. Users in the $n$th group share the $n$th pilot sequence. Their delays are represented by $\tau_{kn}$, $k = 1, 2, \dots, K$. The pilot sequences are chosen from an identity matrix. The $n$th pilot sequence is the $n$th row vector of the identity matrix. If the interference between the users in a group is forced to be zero, the following condition must be satisfied:
	\begin{equation}
	\begin{aligned}
	\tau_{k_1n_1} \leqslant \tau_{k_2n_2},\ \ \mathrm{for}\ n_1 < n_2\ \mathrm{and}\ \forall k_1,\ k_2. 
	\end{aligned}
	\end{equation}
	\label{differentgroup}
\end{theorem}

\begin{IEEEproof}
	The proof is presented in Appendix \ref{proofTheorem1}.
\end{IEEEproof}



Since there is no interference among different user groups, the optimal condition within each user group would be the optimal case for all the users. In the following, we will separate the analysis into two parts: (i) the equally-divided delay scheme within each user group and (ii) the equally-divided delay scheme among different user groups.

\subsection{Equally-divided Delay Scheme within Each User Group}

\begin{figure}[t b]
	\centering
	\includegraphics[width=3in]{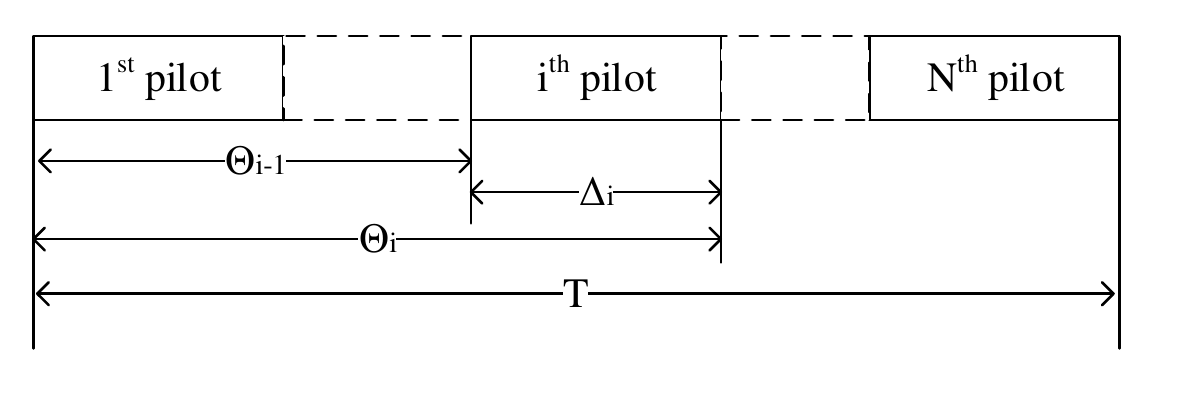}
	\caption{Partition of distinct pilot sequences within the symbol length.}
	\label{equal-divided group}
\end{figure}

In this subsection, we focus on the interference within each group, which is the result of pilot contamination. Without loss of generality, we choose the $i$th group randomly as an example. As is shown in Fig. \ref{equal-divided group}, delays of users sharing the $i$th pilot sequence should be chosen from $\theta_{i - 1}$ and $\theta_{i}$, i.e., $0 \leqslant \theta_{i-1} \leqslant \tau_{1i} < \tau_{2i} < \dots < \tau_{Ki} \leqslant \theta_i < T$. To simplify our notations, we introduce the normalized variables $\delta_k = (\tau_{i(k+1)} - \tau_{ik}) / T,\ k = 1, 2, \dots, K-1$ where $\sum_{k = 1}^{K-1}\delta_k \leqslant (\theta_{i} - \theta_{i-1}) / T = \Delta_i / T \leqslant 1$. For simplicity, we omit the group index in $\delta$s because it will not affect the following discussion. Then, when the identity matrix is employed as the pilot matrix and there is no inter-group interference, $\mathbf{A}$ for Group $i$ can be rewritten as

\begin{equation}
\begin{aligned}
\mathrm{\mathbf{A}} = \left[\begin{smallmatrix}
1\ &1-\delta_1\ &\dots\ &1-\sum_{i = 1}^{K-2}\delta_i\ &1-\sum_{i = 1}^{K-1}\delta_i\\
1-\delta_1\ &1\ &\dots\ &1-\sum_{i = 2}^{K-2}\delta_i\ &1-\sum_{i = 2}^{K-1}\delta_i\\
\vdots\ &\vdots\ &\vdots\ &\vdots\ &\vdots\\
1-\sum_{i = 1}^{K-2}\delta_i\ &1-\sum_{i = 2}^{K-2}\delta_i\ &\dots\ &1\ &1-\delta_{K-1}\\
1-\sum_{i = 1}^{K-1}\delta_i\ &1-\sum_{i = 2}^{K-1}\delta_i\ &\dots\ &1-\delta_{K-1}\ &1
\end{smallmatrix}\right].
\end{aligned}
\label{Asimple}
\end{equation}

The following theorem proves the optimality of the equally-divided delay scheme.

\begin{theorem}
	For the number of cells $2 \leqslant K \leqslant 7$, the MSE of estimated channel matrix will be minimized if the user time delays in Group $i$ are equally divided, i.e.,
	\begin{equation}
	\begin{aligned}
	\delta_j = \frac{\Delta_i}{(K-1)T},\ j = 1,2,\dots, K - 1.
	\end{aligned}
	\end{equation}
	\label{equal_divided_within_group}
\end{theorem}

\begin{IEEEproof}
	The proof is presented in Appendix \ref{proofTheorem2}.
\end{IEEEproof}

We limit $K \geqslant 2$ because the number of variables $\delta$s is $K-1$. If $K = 1$, there is no interference within a user group. The maximum number of cells, $K$, is 7 because there are at most 6 neighboring cells around any specific cell in the hexagonal cell model. We only consider signals from users in any specific cell and its neighboring cells, which means $K$ is less than or equal to 7.\par   

Note that we claim that time delays should be equally divided within the symbol length according to Theorem \ref{equal_divided_within_group}. Considering that time delays can be manipulated at the transmitter side, the time delays of received signals do not necessarily depend on the users' location.\par

Applying Theorem \ref{equal_divided_within_group}, user delays in Group $i$ can be expressed as
\begin{equation}
\begin{aligned}
\tau_{ji} =
\theta_{i-1} + \frac{(j - 1) \Delta_i}{K - 1},\ j = 1, \dots, K, 
\end{aligned}
\label{delayswithingroup}
\end{equation}
where $2 \leqslant K \leqslant 7$, $\Delta_i = \theta_i - \theta_{i-1}$. 

\subsection{Dividing Delays among User Groups}

We have proved that the delays must be equally divided within each group, i.e., $\delta_k = \delta_i^\prime = \frac{\Delta_i}{(K-1)T},\ k = 1, 2, \dots, K-1$, and $K = 2, \dots, 7$. Applying this condition, $\mathbf{A}$ can be written as
\begin{equation}
\begin{aligned}
\mathbf{A} = \left[\begin{smallmatrix}
1\ &1-\delta_i^\prime\ &1-2\delta_i^\prime\ &\dots\ &1-(K-1)\delta_i^\prime\\
1-\delta_i^\prime\ &1\ &1-\delta_i^\prime\ &\dots\ & 1-(K-2)\delta_i^\prime\\
\vdots\ &\vdots\ &\vdots\ &\ddots\ &\vdots\\
1-(K-1)\delta_i^\prime\ &1-(K-2)\delta_i^\prime\ &1-(K-3)\delta_i^\prime\ &\dots\ &1
\end{smallmatrix}\right].
\end{aligned}
\end{equation}

The MSE of users' estimated channels in Group $i$ can be derived analytically. According to Theorem 1 in \cite{bunger2014inverses}, 
\begin{equation}
\begin{aligned}
\mathbf{A}^{-1} = \frac{1}{2\delta_i^\prime}\left[\begin{smallmatrix}
1-\mu\ &-1\ &0\ &\dots\ &0\ &0\ &-\mu\\
-1\ &2\ &-1\ &\dots\ &0\ &0\ &0\\
\vdots\ &\vdots\ &\vdots\ &\ddots\ &\vdots\ &\vdots\ &\vdots\\
0\ &0\ &0\ &\dots\ &-1\ &2\ &-1\\
-\mu\ &0\ &0\ &\dots\ &0\ &-1\ &1-\mu\\
\end{smallmatrix}\right],
\end{aligned}
\end{equation}
where $\mu = \frac{\delta_i^\prime}{(K - 1)\delta_i^\prime - 2}$. Then, the MSE in Group $i$ MSE$_i$ can be written as
\begin{equation}
\begin{aligned}
\mathrm{MSE}_i &= \frac{1}{K\gamma}tr(\mathbf{A}^{-1})
= \frac{1}{K\gamma}\left(\frac{K-1}{\delta_i^\prime} - \frac{1}{(K - 1)\delta_i^\prime - 2}\right)
= \frac{T}{K\gamma} \left(\frac{(K-1)^2}{\Delta_i} + \frac{1}{2T - \Delta_i}\right).
\end{aligned}
\end{equation}

The derivative of MSE$_i$ with respect to $\Delta_i$ is expressed as
\begin{equation}
\begin{aligned}
\frac{\mathrm{d}\ \mathrm{MSE}_i}{\mathrm{d}\Delta_i} = \frac{T}{K\gamma}\left(-\frac{(K - 1)^2}{\Delta_i^2} + \frac{1}{(2T - \Delta_i)^2}\right)< 0,
\end{aligned}
\label{43}
\end{equation}
for $2 \leqslant K \leqslant 7,\ \Delta_i\in [0, T]$, and $\sum_{i = 1}^{N}\Delta_i = T$. Eq. (\ref{43}) shows that MSE$_i$ is a monotonic function of $\Delta_i$. Since there is no interference between different user groups, the average of MSE is,
\begin{equation}
\begin{aligned}
\mathrm{MSE} = \frac{1}{N}\sum_{i = 1}^{N}\mathrm{MSE}_i \geqslant \sqrt[N]{\prod_{i = 1}^{N}\mathrm{MSE}_i},
\end{aligned}
\end{equation}
where the equality is achieved when all MSE$_i$s are equal. Remember that MSE$_i$ is a monotonic function of $\Delta_i$. As a result, the fact that all MSE$_i$s are equal leads to the conclusion that all $\Delta_i$s are equal to $T/N$, which is the equally-divided scheme among user groups. Employing our equally-divided scheme, the minimum MSE that the asynchronous channel training method could achieve using the ZF estimator in the $K$-cell $N$-user system can be expressed as
\begin{equation}
\begin{aligned}
\mathrm{MSE} = \frac{N}{K\gamma}\left((K-1)^2 + \frac{1}{2N - 1}\right).
\end{aligned}
\end{equation}

\begin{figure}[t b]
	\centering
	\includegraphics[width=3in]{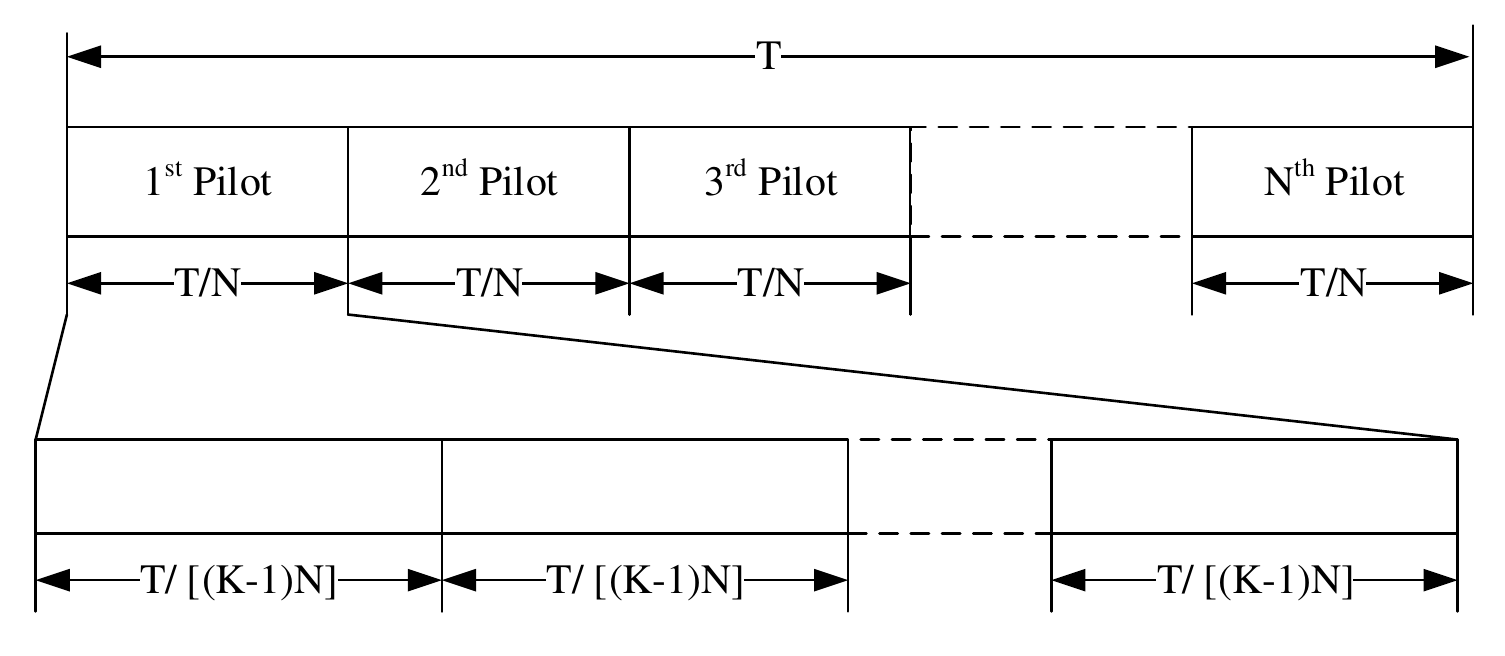}
	\caption{Distribution of user delays in equally-divided delay scheme for $K$-cell $N$-user scenarios.}
	\label{distribution}
\end{figure}

The equally-divided delay scheme is shown in Fig. \ref{distribution}. To sum up, delays of users employing the $i$th pilot sequence belong to the range $[(i - 1)T/N, iT/N - T]$, $i = 1, 2, \dots, N$. Within the range, the $j$th largest delay can be expressed as
\begin{equation}
\tau_{ji} = \left(i - 1 + \frac{j - 1}{K - 1}\right)\frac{T}{N},\ j = 1, \dots, K.
\label{delaydistribution}
\end{equation} 

Note that the equally-divided delay scheme is the optimal solution to minimize MSE based on the following preconditions: (i) Theorem \ref{differentgroup} is satisfied, i.e., there is no inter-group interference; (ii) ZF estimation is employed in the asynchronous channel training approach; (iii) The identity matrix is used as the pilot matrix. 

\section{Performance Analysis}
After channel training, data will be transmitted from UE to BS in uplink. As is shown in \cite{ganji,verdu1989capacity}, the multiuser interference can be mitigated via asynchronous transmission and sampling diversity. However, we avoid using oversampling during data transmission in order to demonstrate the effectiveness of our asynchronous channel training scheme to mitigate the pilot contamination and have a fair comparison with the synchronous case. In fact, the uplink achievable rate will be higher if oversampling is taken into consideration for the asynchronous scenario. We assume that the channel estimation and data transmission phases share the same SNR at receiver, $\gamma$, and the channel does not change during the data transmission block. The signal sample, synchronous to UE$_{kn}$, at time $i$ $(1 < i < L)$ after the sampler which is synchronous to UE$_{kn}$ can be written as
\begin{equation}
\begin{aligned}
\mathbf{y}_{kn}[i] = \sqrt{\gamma}\mathbf{h}_{kn}x_{kn}[i] + \sqrt{\gamma}\underset{(k^\prime, n^\prime) \neq (k, n)}{\sum_{k^\prime = 1}^{K}\sum_{n^\prime = 1}^{N}} \mathbf{h}_{k^\prime n^\prime} \mathbf{r}_{k^\prime n^\prime}^{kn} \mathbf{x}_{k^\prime n^\prime} + \mathbf{n}_{k^\prime n^\prime}[i],
\end{aligned}
\end{equation}
where $\mathbf{x}_{k^\prime n^\prime}$ is the $L\times 1$ column vector denoting the $L$ data symbols from UE$_{k^\prime n^\prime}$, $\mathbf{r}_{k^\prime n^\prime}^{kn}$ is a $1 \times L$ row vector representing coefficients of asynchronous transmission. If the rectangular pulse shape is used and $\tau_{kn} > \tau_{k^\prime n^\prime}$, $\mathbf{r}_{k^\prime n^\prime}^{kn}$ is given as
\begin{equation}
\begin{aligned}
\mathbf{r}_{k^\prime n^\prime}^{kn} = [\underbrace{0,\dots, 0}_{i-2}, \rho_{k^\prime n^\prime}^{kn}, \rho^{k^\prime n^\prime}_{kn}, 0,\dots, 0].
\end{aligned}
\end{equation}
If $\tau_{kn} < \tau_{k^\prime n^\prime}$, $\mathbf{r}_{k^\prime n^\prime}^{kn}$ is given as
\begin{equation}
\begin{aligned}
	\mathbf{r}_{k^\prime n^\prime}^{kn} = [\underbrace{0,\dots, 0}_{i-1}, \rho^{k^\prime n^\prime}_{kn}, \rho_{k^\prime n^\prime}^{kn}, 0,\dots, 0].
\end{aligned}
\end{equation}
If the two signals are synchronous, i.e., $\tau_{kn} = \tau_{k^\prime n^\prime}$,
\begin{equation}
\begin{aligned}
\mathbf{r}_{k^\prime n^\prime}^{kn} = [\underbrace{0,\dots, 0}_{i-1}, 1, 0,\dots, 0].
\end{aligned}
\end{equation} 

Using MRC, the detected symbol $\hat{x}_{kn}[i]$ can be written as
\begin{equation}
\begin{aligned}
&\hat{x}_{kn}[i] = \hat{\mathbf{h}}_{kn}^H \mathbf{y}_{kn}[i]\\
& = \sqrt{\gamma}\hat{\mathbf{h}}_{kn}^H \mathbf{h}_{kn}x_{kn}[i] + \sqrt{\gamma}\underset{(k^\prime, n^\prime) \neq (k, n)}{\sum_{k^\prime = 1}^{K}\sum_{n^\prime = 1}^{N}} \hat{\mathbf{h}}_{kn}^H
\mathbf{h}_{k^\prime n^\prime} \mathbf{r}_{k^\prime n^\prime}^{kn} \mathbf{x}_{k^\prime n^\prime} + \hat{\mathbf{h}}_{kn}^H n_{k^\prime n^\prime}[i]\\
&= \sqrt{\gamma}\hat{\mathbf{h}}_{kn}^H \hat{\mathbf{h}}_{kn}x_{kn}[i] + \sqrt{\gamma}\hat{\mathbf{h}}_{kn}^H (\mathbf{h}_{kn} - \hat{\mathbf{h}}_{kn}^H) x_{kn}[i] + \sqrt{\gamma}\underset{(k^\prime, n^\prime) \neq (k, n)}{\sum_{k^\prime = 1}^{K}\sum_{n^\prime = 1}^{N}} \hat{\mathbf{h}}_{kn}^H
\mathbf{h}_{k^\prime n^\prime} \mathbf{r}_{k^\prime n^\prime}^{kn} \mathbf{x}_{k^\prime n^\prime} + \hat{\mathbf{h}}_{kn}^H n_{k^\prime n^\prime}[i].
\end{aligned}
\end{equation} 

With $x_{kn}[i] \sim \mathcal{CN}(0, 1)$, $k = 1,2,\dots, K$, $n = 1,2,\dots,N$, for any $i$, the instantaneous SINR of UE$_{kn}$ can be expressed as \cite{antonios2017icc}
\begin{equation}
\begin{aligned}
\mathrm{SINR}_{kn} = \frac{\gamma ||\hat{\mathbf{h}}_{kn}||^4}{\gamma ||\hat{\mathbf{h}}_{kn}^H (\mathbf{h}_{kn} - \hat{\mathbf{h}}_{kn}^H)||^2 + \gamma \underset{(k^\prime, n^\prime) \neq (k, n)}{\sum_{k^\prime = 1}^{K}\sum_{n^\prime = 1}^{N}} ||\hat{\mathbf{h}}_{kn}^H	\mathbf{h}_{k^\prime n^\prime}||^2 ||\mathbf{r}_{k^\prime n^\prime}^{kn}||^2 + ||\hat{\mathbf{h}}_{kn}||^2}.
\end{aligned}
\end{equation}

Using the above SINR, the instantaneous achievable rate of UE$_{kn}$ is
\begin{equation}
\begin{aligned}
R_{kn} = \mathrm{log}_2\left(1 + \mathrm{SINR}_{kn}\right).
\end{aligned}
\end{equation}
 
The average instantaneous achievable rate per UE is calculated as
\begin{equation}
\begin{aligned}
R_{\mathrm{perUE}} = \frac{1}{KN} \sum_{k = 1}^{K}\sum_{n = 1}^{N} \mathrm{log}_2\left(1 + \mathrm{SINR}_{kn}\right).
\end{aligned}
\end{equation}

The average achievable rate per UE is computed by averaging out at random channels, i.e.,
\begin{equation}
\begin{aligned}
R_{\mathrm{average}} = \mathbb{E}\left[\frac{1}{KN} \sum_{k = 1}^{K}\sum_{n = 1}^{N} \mathrm{log}_2\left(1 + \mathrm{SINR}_{kn}\right)\right],
\end{aligned}
\label{average_rate}
\end{equation}
which is used as the criterion for performance comparison in the next section. 

\section{Simulation Results}
In this section, we provide simulation results for the asynchronous channel training scheme. In our simulation, we use the conventional synchronous system suffering from pilot contamination as benchmark to compare with our asynchronous system. In the legend of each figure, ``asynchronous w/ oversampling'', ``synchronous w/o oversampling'', and ``asynchronous w/o oversampling'' represent our proposed asynchronous channel training scheme, the conventional synchronous channel training scheme in the synchronous scenario, and the conventional synchronous channel training scheme employed in the asynchronous scenario, respectively. 

The symbol length $T$ is normalized, i.e., $T = 1$. The default number of antennas at BS is 100 and the identity matrix is employed as the pilot matrix unless mentioned otherwise. Each user has a single antenna. For the case with random delays, the time delay of each received signal $\tau$ is uniformly distributed between $0$ and $1$. 

\begin{figure}[t b]
	\centering
	\includegraphics[width=3.5in]{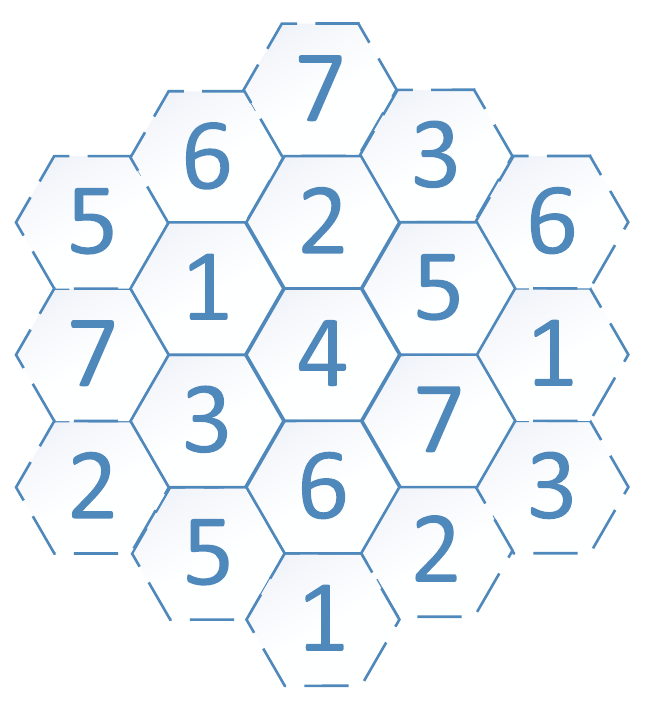}
	\caption{Topology of grid hexagonal cells.}
	\label{topology}
\end{figure}

In simulation, the rectangular pulse shape $s(t)=u(t)-u(t-1)$ is employed and $\int_{0}^{1}s^2(t)dt=1$. We assume that the channel matrix $\mathbf{H}$ is distributed as a zero-mean complex Gaussian matrix. For each BS, the channel matrix entries of its serving UEs are unit-variance while those of other cells have a variance of 0.5. The number of orthogonal pilot sequences is equal to the number of users in the cell, $N$. In simulation, we adopt two pilot matrices, the $N$-by-$N$ normalized DFT matrix and the identity matrix. The achievable rate of the random case is calculated as the average achievable rate with each delay configuration. In our simulation, all channel matrices are estimated by LMMSE estimator because LMMSE is widely used in practice. For the synchronous case, $\mathbf{R}$ is substituted by Eq. (\ref{R_syn}) and additive noise is white Gaussian, i.e., $\mathbf{R_{NN}}=\mathbf{I}_{KN\times KN}$. 

Furthermore, we calculate the average achievable rate per UE as the performance metric, using Eq. (\ref{average_rate}). Take the 7-cell scenario, $K = 7$, as an example, which is shown as Fig. \ref{topology}. We consider the average achievable rate of UEs in the central 7 cells with solid edge lines. The index of each cell represents the index of time delay parameter sets. The UEs in the cells with the same index share the same set of time delays. We employ the wraparound method to eliminate the edge effect when calculating the achievable rate of UEs in edge cells.

\begin{figure}[t b]
	\centering
	\includegraphics[width=3.5in]{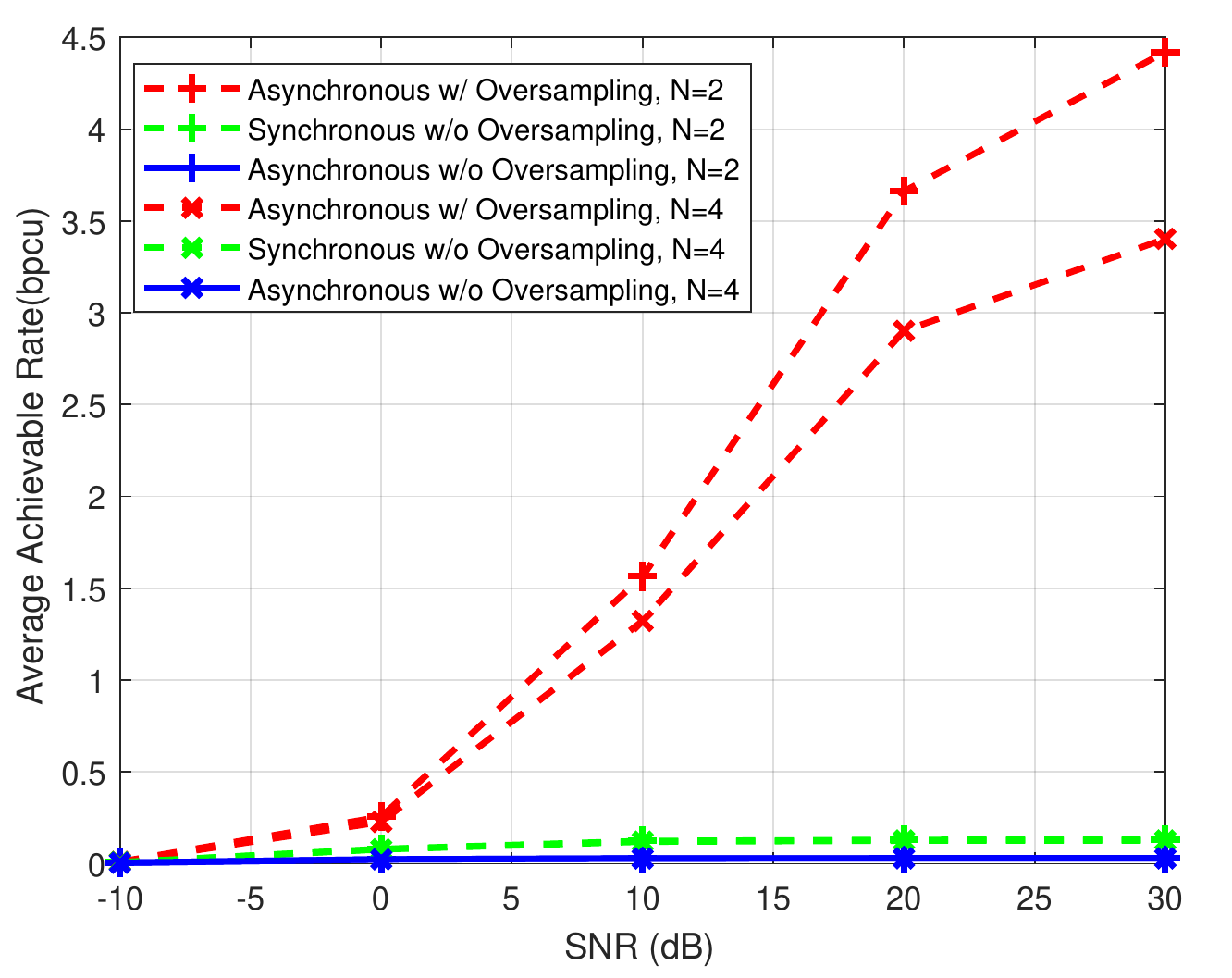}
	\caption{Performance of difference channel training schemes as a function of SNR for different number of users in each cell ($N = 2, 4$), with 7 cells in total, using $M = 100$ receive antennas.}
	\label{rate_vs_snr_N}
\end{figure}

\subsection{Rate vs. SNR for Different Number of Users}
It is shown in Fig. \ref{rate_vs_snr_N} that the average achievable rate of the asynchronous channel training scheme with oversampling increases while the performance of the synchronous and asynchronous cases without oversampling saturate as the receive SNR increases. It is because the channel matrices are contaminated by inter-cell interference in uplink training using the channel estimation methods without oversampling in both synchronous and asynchronous cases. In contrast, our asynchronous channel training scheme takes advantage of sampling diversity to reduce the negative effect of pilot contamination. Furthermore, the system with less users achieves better performance because it causes less inter-user interference.

\begin{figure}[t b]
	\centering
	\includegraphics[width=3.5in]{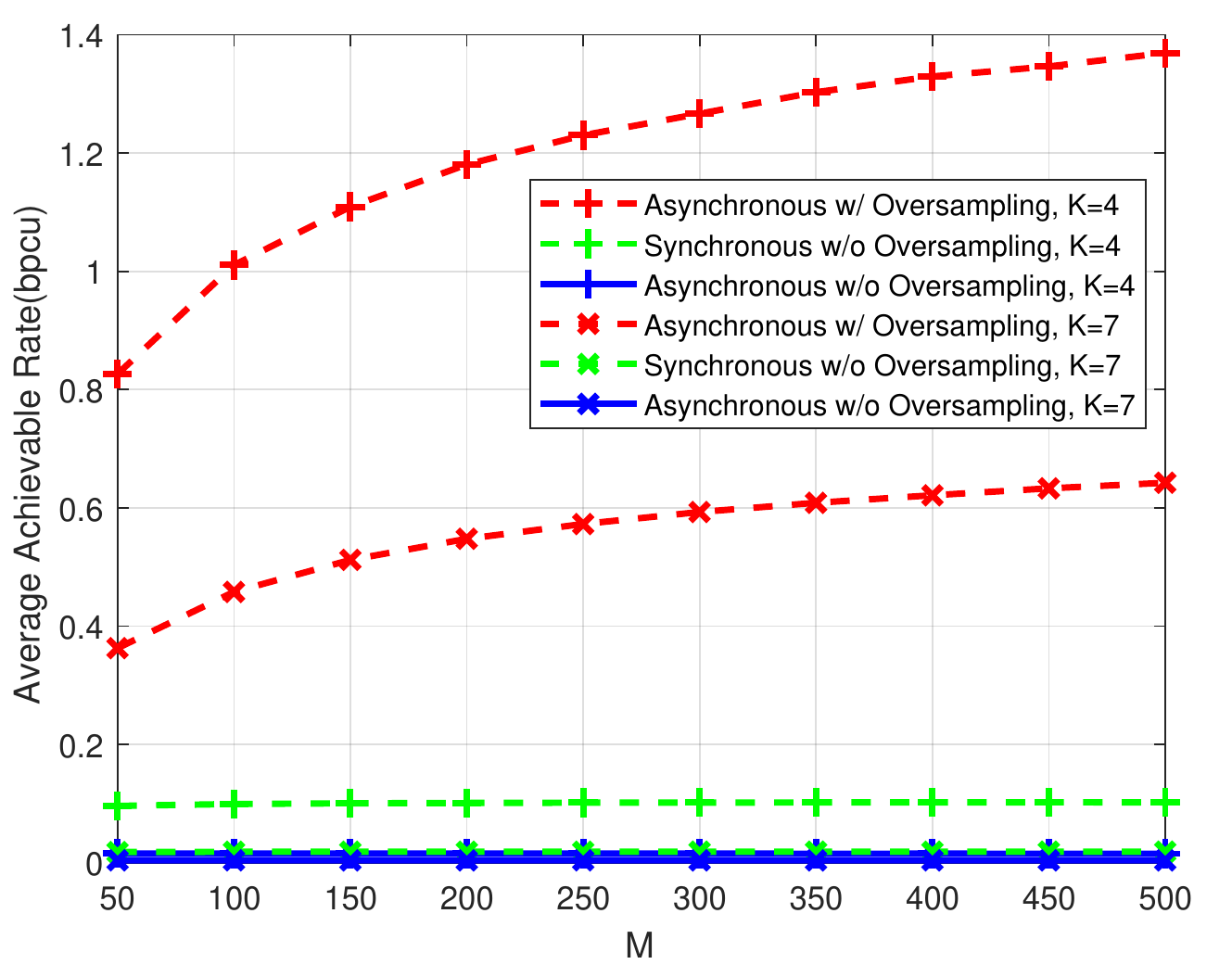}
	\caption{Performance of our asynchronous channel training scheme and the synchronous training scheme (with and without oversampling) for different number of cells $K=4,7$ with 3 users in each cell.}
	\label{rate_vs_M_K}
\end{figure}

\subsection{Rate vs. M for Different Number of Cells}
Fig. \ref{rate_vs_M_K} shows the average achievable rate as a function of the antenna number for different number of cells, $K=4,7$, with receive SNR $\gamma = 20$dB. We assume that there are 3 users in each cell. As expected, for the asynchronous case, the average achievable rate is improved by increasing $M$. The performance of synchronous and asynchronous cases without oversampling do not change with $M$ because the pilot contamination dominants the performance at high SNR regime. This phenomenon verifies the widely accepted conclusion that the system performance saturates in synchronous multi-cell multi-user massive MIMO systems \cite{jose2011pilot}. Moreover, the asynchronous channel has a negative influence on the channel estimation accuracy if a system designed for the synchronous case is used. It is demonstrated that the performance of the asynchronous channel training scheme is better than that of the synchronous scheme. In addition, the estimation performance of both the synchronous scheme and the asynchronous scheme deteriorates as the number of cells, $K$, increases since more cells lead to more severe pilot contamination.

\begin{figure}[t b]
	\centering
	\includegraphics[width=3.5in]{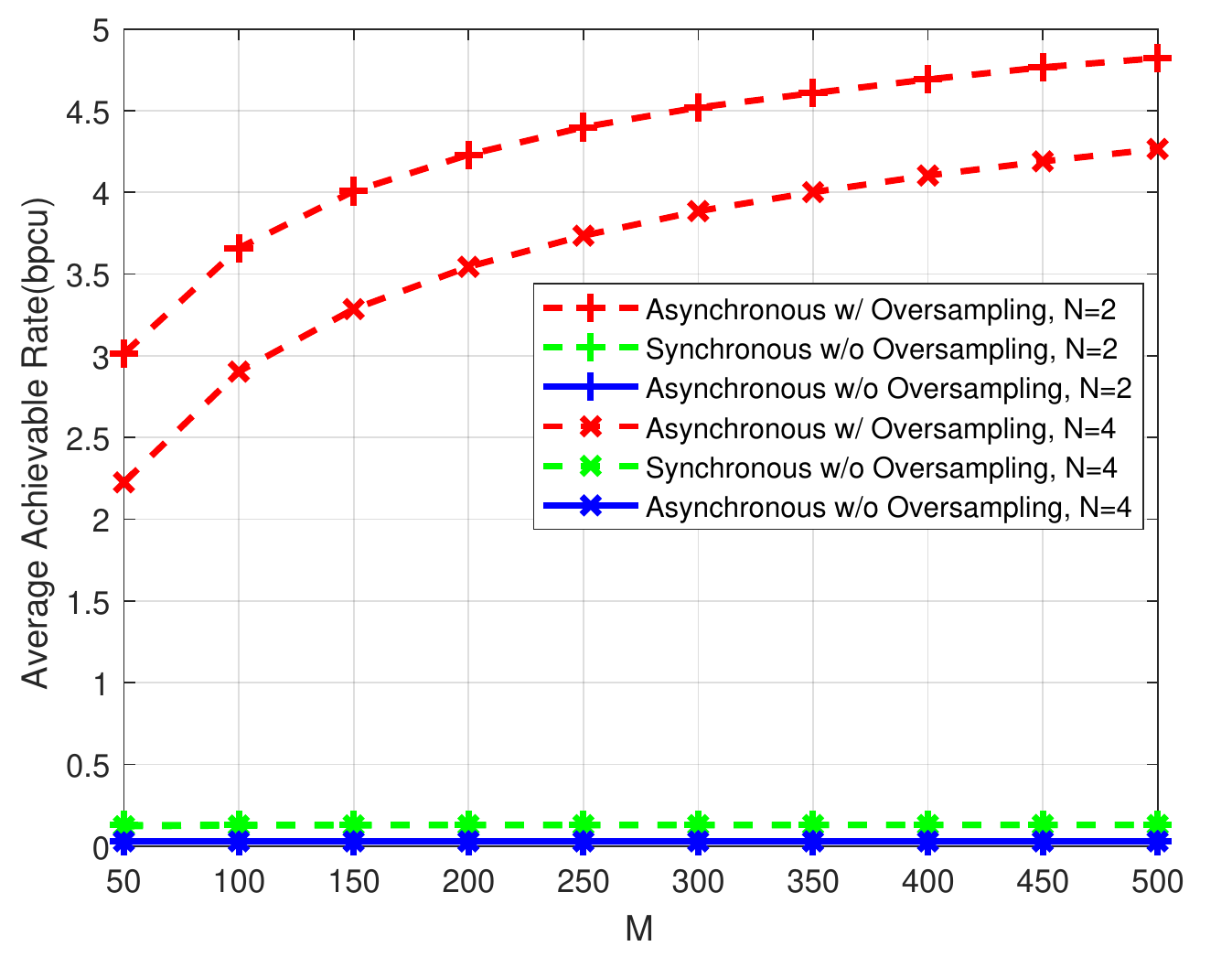}
	\caption{Performance of our asynchronous channel training scheme and the synchronous training scheme (with and without oversampling) for different number of users in each cell $N=2,4$ with 7 cells in total.}
	\label{rate_vs_M_N}
\end{figure}

\subsection{Rate vs. M for Different Number of Users}  
Fig. \ref{rate_vs_M_N} shows the average achievable rate as a function of the antenna number for different number of users in each cell with received SNR $\gamma = 20$dB. The total number of cells is 7. It is obvious that our asynchronous channel training scheme provides performance improvement compared with the conventional synchronous system with pilot contamination. Moreover, the average achievable rate increases with the number of antennas because larger antenna array provides higher array gain.

\begin{figure}[t b]
	\centering
	\includegraphics[width=3.5in]{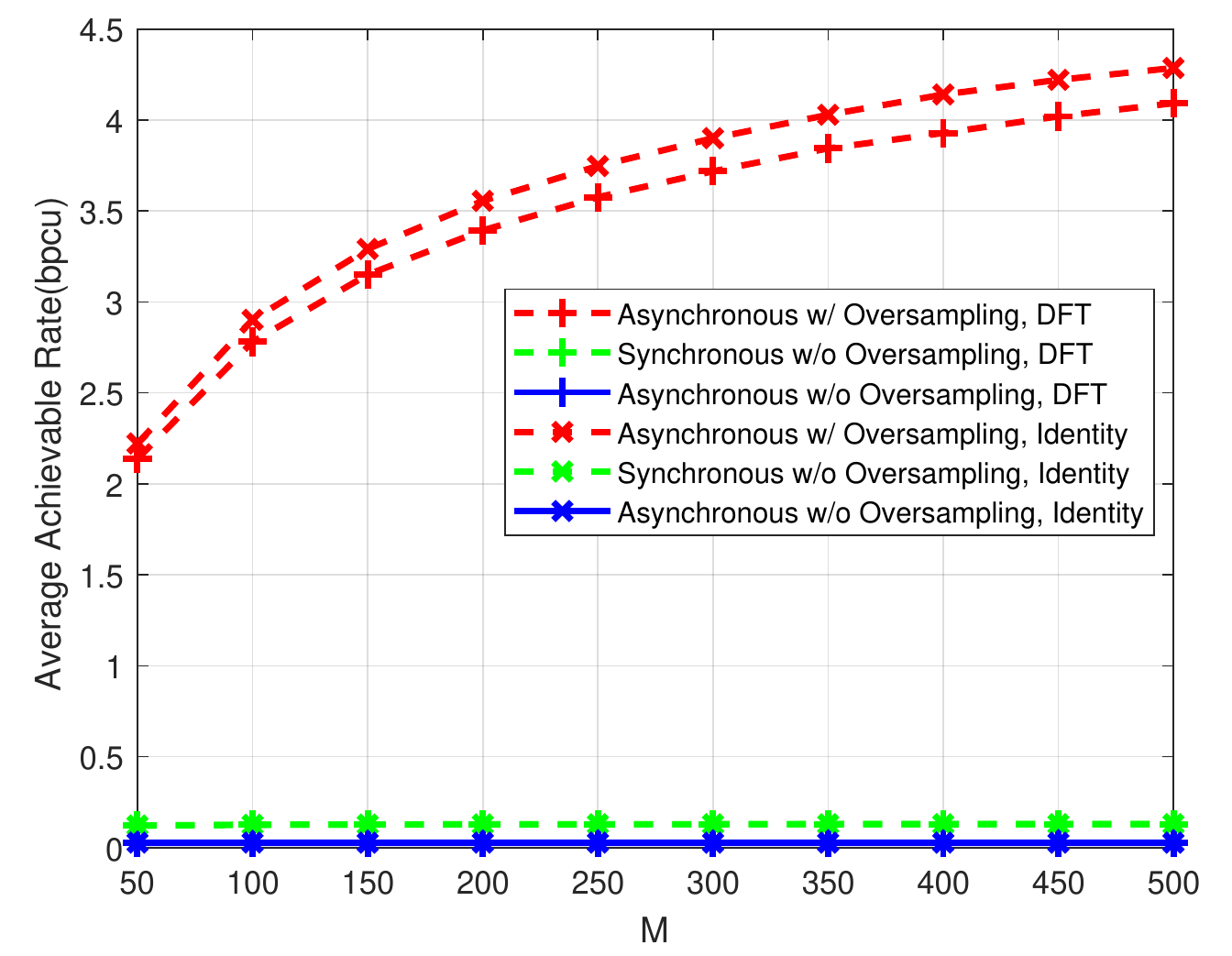}
	\caption{Performance of our asynchronous channel training scheme and the synchronous scheme (with and without oversampling) for different pilot matrices with receive SNR $\gamma = 20$dB. The total number of cells is 7 with 3 users located in each cell.}
	\label{rate_vs_M_pilotmode}
\end{figure}

\subsection{Rate vs. M for Different Kinds of Pilot Sequences}
Fig. \ref{rate_vs_M_pilotmode} shows the average achievable rate as a function of the antenna number for different pilot matrices, i.e., identity matrix and normalized DFT matrix. There are 7 cells with 3 users located in each cell. As shown in Fig. \ref{rate_vs_M_pilotmode}, the choice of the pilot matrix, as long as it is orthogonal, has almost no effect on the estimation error in the synchronous and asynchronous cases without oversampling. In the asynchronous case with oversampling, the system using identity matrix achieves better performance than DFT matrix because each symbol in the DFT matrix has the same power. On the contrary, for each pilot sequence in the identity matrix, only one symbol has a non-zero power. This feature is beneficial to relieve ISI and inter-user interference to obtain more accurate channel matrices.

\begin{figure}[t b]
	\centering
	\includegraphics[width=3.5in]{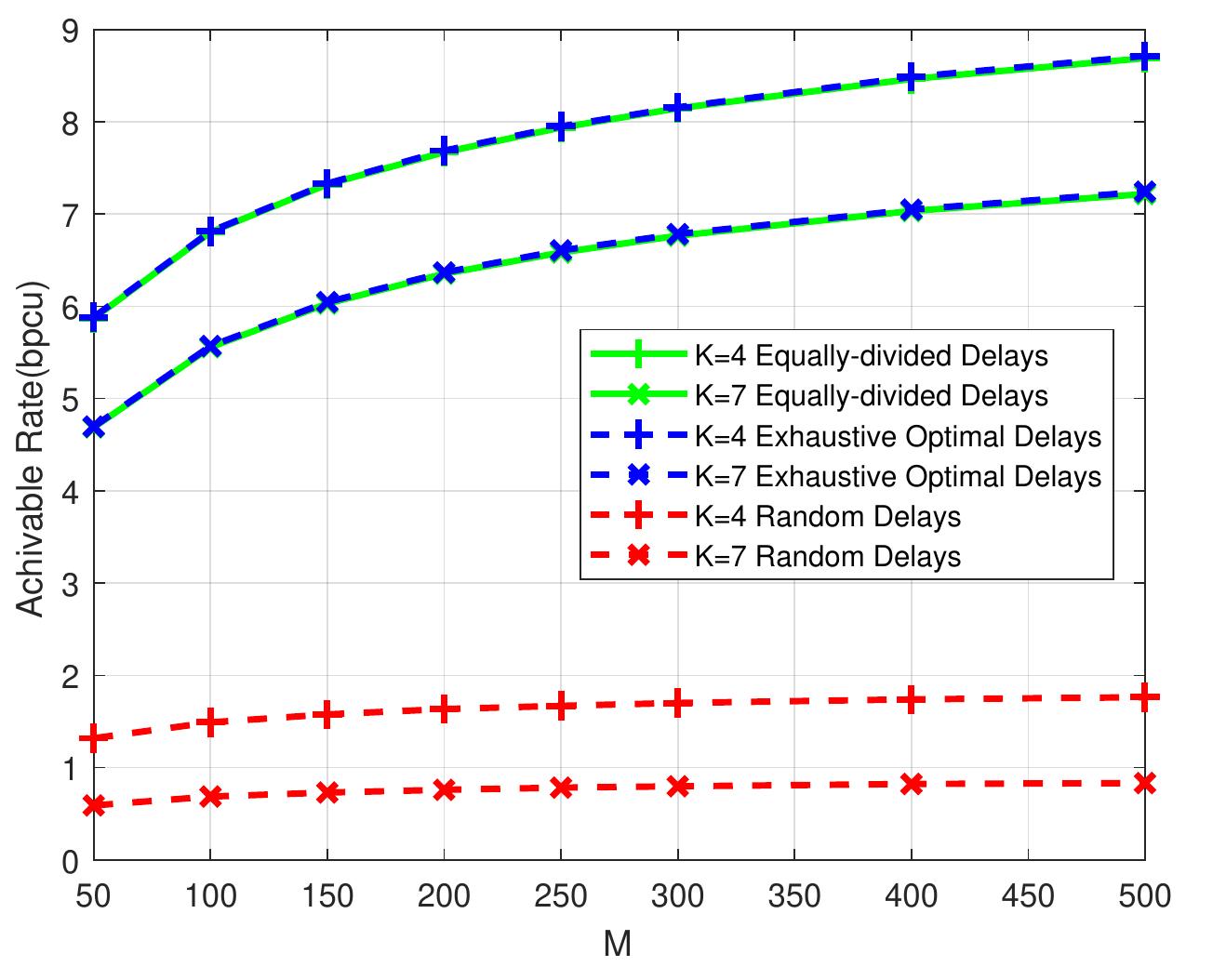}
	\caption{Performance comparison of asynchronous channel training scheme using random, exhaustively optimal, and equally-divided delays.}
	\label{rate_comparisonMU}
\end{figure}

\subsection{Rate vs. M for Different Delay Schemes}
In Fig. \ref{rate_comparisonMU}, the average achievable rate performance for random, exhaustively optimal, and equally-divided delays are compared with receive SNR $\gamma = 20$dB. We consider two multi-cell single-user scenarios, including the 4-cell case and the 7-cell case. We provide the average achievable rates for the globally optimal delay scheme found via exhaustive search as benchmark. In exhaustive search, we set the delay of one user as 0 and traverse delay variables of all the other users from 0 to 1 with interval 0.05. The exhaustively optimal delay scheme is the one corresponding to the highest achievable rate. Simulation results indicate that the optimal time delays derived by exhaustive search are almost equally divided. Furthermore, the equally-divided delay scheme provides similar performance as the optimal scheme. Note that we have analytically proved that the equally-divided delay scheme is the optimal solution for the ZF estimator, which, as a result, should be the optimal case for the LMMSE estimator at high SNRs as well.

%

\subsection{Summary}
In summary, our asynchronous channel training method can provide performance gain. The performance of the asynchronous channel training scheme deteriorates as the cell number increases. The number of users in each cell and the number of receive antennas have little effect on the performance. Better performance is achieved if the identity matrix is employed as the pilot matrix.

\section{Conclusion}
In this paper, we proposed a novel asynchronous channel training method. Based on this method, LMMSE and ZF estimators were developed. Moreover, the equally-divided delay scheme was designed to optimize the MSE performance of the ZF estimator. We proved that the user delays have to be equally divided within the symbol length in order to minimize the MSE of the ZF estimator if the identity matrix is used as the pilot matrix and there is no interference among users sharing the same pilot sequence. It was shown that the asynchronous channel training scheme could achieve better performance compared to the synchronous scheme suffering from pilot contamination.


%

\appendices
\section{Proof of Theorem \ref{differentgroup}}
\label{proofTheorem1}

\begin{figure}[t b]
	\centering
	\includegraphics[width=3in]{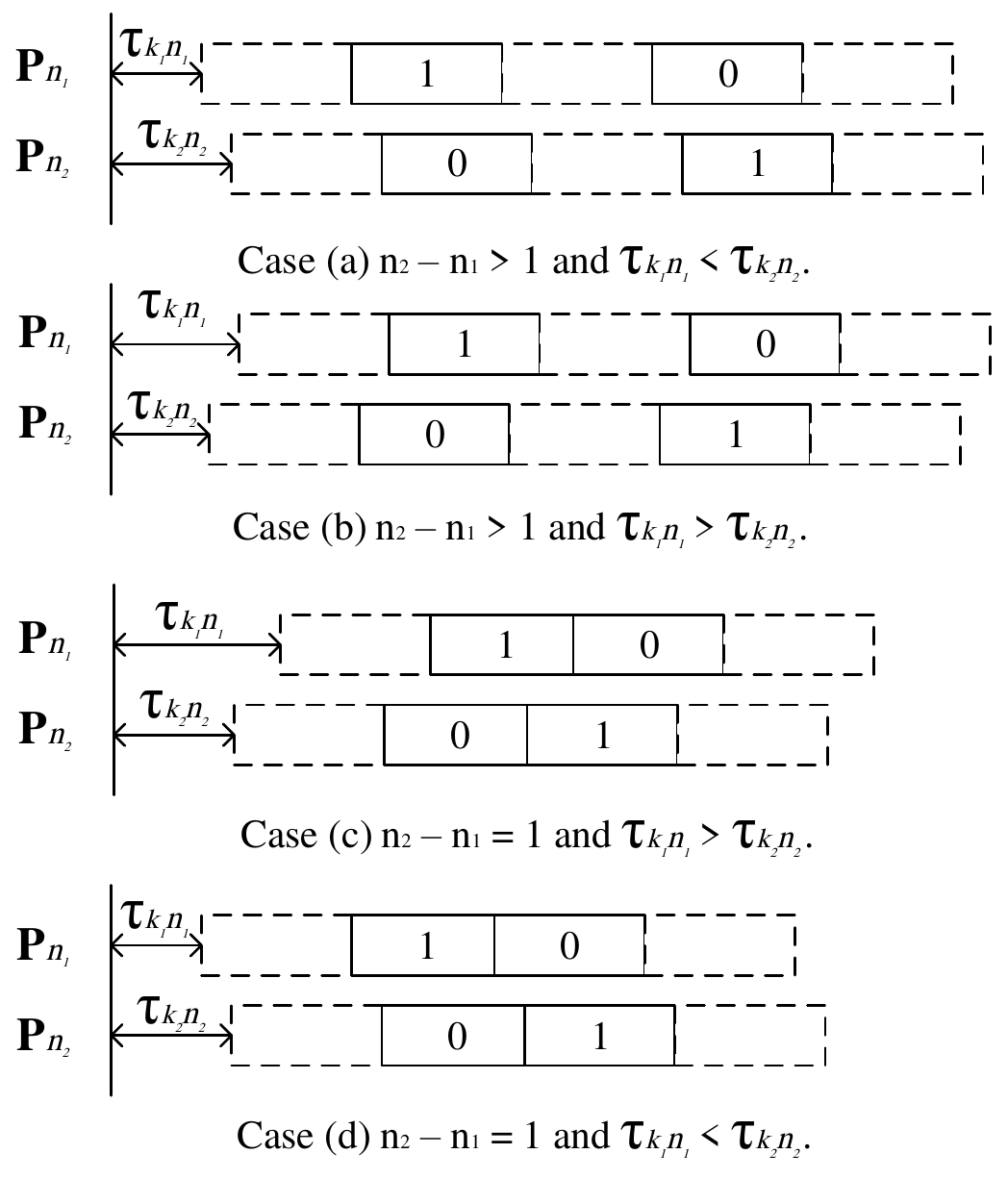}
	\caption{Four possible cases for users employing different pilot sequences.}
	\label{theorem1}
\end{figure}

\begin{IEEEproof}
	In Eq. (\ref{A}), the entries $\mathbf{P}_{n_1}\mathbf{R}^{k_1n_1}_{k_2n_2}\mathbf{P}^H_{n_2}$ where $n_1 \neq n_2$ stand for the inter-group interference. Because $\mathbf{P}_{n}$ is the $n$th row of an $N \times N$ identity matrix, $\mathrm{\mathbf{P}}_{n_1}\mathrm{\mathbf{R}}^{k_1n_1}_{k_2n_2}\mathrm{\mathbf{P}}^H_{n_2} = \mathrm{\mathbf{R}}^{k_1n_1}_{k_2n_2}(n_1, n_2)$, i.e., the $(n_1, n_2)$th entry of $\mathrm{\mathbf{R}}^{k_1n_1}_{k_2n_2}$. We assume that $n_2 > n_1$ since the proof will be the same if $n_2 < n_1$. As all the delays will not exceed the symbol length $T$, there will be no interference if $n_2 - n_1 > 1$, which is depicted by Cases (a) and (b) in Fig. \ref{theorem1}. It is also easy to show these cases mathematically: the $(n_1, n_2)$th entry in Eqs. (\ref{lowertriangle}) and (\ref{uppertriangle}) will always be zero if $n_2 - n_1 > 1$. Therefore, we only need to worry about the interference when $n_2 - n_1 = 1$.
	
	If $\tau_{k_1n_1} > \tau_{k_2n_2}$ ($\forall k_1,\ k_2$) which is shown in Case (c) of Fig. \ref{theorem1}, the two symbol 1s in $\mathbf{P}_{n_1}$ and $\mathbf{P}_{n_2}$ overlap partially, leading to interference. Mathematically, all ($n_1, n_1 + 1$) entries are non-zero in Eq. (\ref{uppertriangle}). On the contrary, if $\tau_{k_1n_1} \leqslant \tau_{k_2n_2}$, all ($n_1, n_1 + 1$) entries are zero in Eq. (\ref{lowertriangle}). This is shown in Case (d) of Fig. \ref{theorem1}. Therefore, the condition $\tau_{k_1n_1} \leqslant \tau_{k_2n_2}$ has to be satisfied if the inter-group interference is forced to be zero. The proof is complete.  
\end{IEEEproof}

\section{Proof of Theorem \ref{equal_divided_within_group}}
\label{proofTheorem2}
To prove Theorem \ref{equal_divided_within_group}, we need to present some lemmas first.

\begin{lemma}
	$f(s) = tr[((1 - s)\mathbf{A} + s \mathbf{JAJ})^{-1}]$ is strictly convex for $s\in (0, 1)$ if $\mathbf{A} \neq \mathbf{JAJ}$.
	\label{convex}
\end{lemma}

\begin{IEEEproof}
	$f(s)$ can be written as
	\begin{equation}
	\begin{aligned}
	f(s) &= tr[((1 - s)\mathbf{A} + s \mathbf{JAJ})^{-1}] 
	= tr[\mathbf{A}^{-1/2}\left(\mathbf{I} + s\mathbf{A}^{-1/2}(\mathbf{JAJ} - \mathbf{A})\mathbf{A}^{-1/2}\right)^{-1}\mathbf{A}^{-1/2}]\\
	& \stackrel{(a)}{=} tr[\mathbf{A}^{-1} (\mathbf{I} + s\mathbf{A}^{-1/2}(\mathbf{JAJ} - \mathbf{A})\mathbf{A}^{-1/2})^{-1}]
	= tr[\mathbf{A}^{-1} (\mathbf{I} + s(\mathbf{A}^{-1/2}\mathbf{JAJ}\mathbf{A}^{-1/2} - \mathbf{I}))^{-1}]\\
	& \stackrel{(b)}{=} tr[\mathbf{A}^{-1} \mathbf{Q}(\mathbf{I} + s(\mathbf{D} - \mathbf{I}))^{-1}\mathbf{Q}^T]
	\stackrel{(c)}{=} tr[\mathbf{Q}^{T}\mathbf{A}^{-1}\mathbf{Q}(\mathbf{I} + s(\mathbf{D} - \mathbf{I}))^{-1}]\\
	& = \sum_{i = 1}^{n} (\mathbf{Q}^T\mathbf{A}^{-1}\mathbf{Q})_{ii} (1 - s + s\lambda_i)^{-1},
	\end{aligned}
	\end{equation}
	where (a) and (c) are obtained via the cyclic property of trace operation, (b) is derived using the eigenvalue decomposition $\mathbf{A}^{-1/2}\mathbf{JAJ}\mathbf{A}^{-1/2} = \mathbf{QDQ}^T$ and exchanging $\mathbf{I}$ as $\mathbf{QQ}^T$, $\lambda_i$ is the $i$th diagonal element of $\mathbf{D}$, and $(\mathbf{Q}^T\mathbf{A}^{-1}\mathbf{Q})_{ii}$ stands for the $(i,i)$th entry of $\mathbf{Q}^T\mathbf{A}^{-1}\mathbf{Q}$. Because $\mathbf{Q}^T\mathbf{A}^{-1}\mathbf{Q}$ is a symmetric positive definite matrix, all its diagonal entries are positive. Then,
	\begin{equation}
	\begin{aligned}
	\frac{\mathrm{d}f}{\mathrm{d}s} &= \sum_{i = 1}^{n} (\mathbf{Q}^T\mathbf{A}^{-1}\mathbf{Q})_{ii} \frac{1 - \lambda_i}{(1 - s + s\lambda_i)^2},
	\end{aligned}
	\end{equation}
	\begin{equation}
	\begin{aligned}
	\frac{\mathrm{d}^2f}{\mathrm{d}s^2} = \sum_{i = 1}^{n} (\mathbf{Q}^T\mathbf{A}^{-1}\mathbf{Q})_{ii} \frac{2(1 - \lambda_i)^2}{(1 - s + s\lambda_i)^3}.
	\end{aligned}
	\end{equation}
	
	Since $\mathbf{A}^{-1/2}\mathbf{JAJ}\mathbf{A}^{-1/2}$ is positive definite, $\lambda_i > 0$. If $\mathbf{A} \neq \mathbf{JAJ}$, $\mathbf{A}^{-1/2}\mathbf{JAJ}\mathbf{A}^{-1/2} \neq \mathbf{I}$ which means that $\exists \lambda_i$, $\lambda_i \neq 1$. As a result, $\mathrm{d}^2f/\mathrm{d}s^2 > 0$ for $s \in (0, 1)$. Therefore, $f(s)$, $s \in (0, 1)$, is strictly convex if $\mathbf{A} \neq \mathbf{JAJ}$. Proof is complete.
\end{IEEEproof}

\begin{lemma}
	In order to minimize the MSE of estimated channels for users using the same pilot sequence, the delay differences should be symmetrically equal, i.e.,
	\begin{equation}
	\begin{aligned}
	\delta_i = \delta_{K - i},\ i = 1,2,\dots,K - 1.
	\end{aligned}
	\end{equation}
	\label{symmetricEq}
\end{lemma}

\begin{IEEEproof}
	Remember that $\mathbf{J}$ is the symmetric elementary matrix which has ones along the secondary diagonal and zeros elsewhere and $\mathbf{JJ} = \mathbf{I}$. Then, $\mathbf{JAJ}$ can be calculated as
		\begin{equation}
		\begin{aligned}
		\mathrm{\mathbf{JAJ}} = \left[\begin{smallmatrix}
		1\ &1-\delta_{K - 1}\ &\dots\ &1-\sum_{i = 2}^{K-1}\delta_i\ &1-\sum_{i = 1}^{K-1}\delta_i\\
		1-\delta_{K - 1}\ &1\ &\dots\ &1-\sum_{i = 2}^{K-2}\delta_i\ &1-\sum_{i = 1}^{K-2}\delta_i\\
		\vdots\ &\vdots\ &\vdots\ &\vdots\ &\vdots\\
		1-\sum_{i = 2}^{K-1}\delta_i\ &1-\sum_{i = 2}^{K-2}\delta_i\ &\dots\ &1\ &1-\delta_{1}\\
		1-\sum_{i = 1}^{K-1}\delta_i\ &1-\sum_{i = 1}^{K-2}\delta_i\ &\dots\ &1-\delta_{1}\ &1
		\end{smallmatrix}\right].
		\end{aligned}
		\end{equation}
		
		Note that $\mathbf{JAJ}$ is obtained if we exchange the locations of every $\delta_i$ and $\delta_{K - i}$, $i = 1, 2, \dots, \lfloor K/2\rfloor$, in $\mathbf{A}$. Considering the function $f(s) = tr[((1 - s)\mathbf{A} + s \mathbf{JAJ})^{-1}]$, it is easy to show that
		\begin{equation}
		\begin{aligned}
		f(0) = tr(\mathbf{A}^{-1}) = tr(\mathbf{JA}^{-1}\mathbf{J}) = tr((\mathbf{JAJ})^{-1})  = f(1).
		\end{aligned}
		\end{equation}
		
		According to Lemma \ref{convex}, $f(s)$ is a strictly convex function, i.e.,
		\begin{equation}
		\begin{aligned}
		f(0) &= f(1) = tr(\mathbf{A}^{-1}) > tr(((1 - s)\mathbf{A} + s \mathbf{JAJ})^{-1}), 
		\end{aligned}
		\end{equation}
		where $s \in (0, 1)$. Moreover, $f(s)$ is symmetric with respect to $s = 1/2$ because 
		
		\begin{equation}
		\begin{aligned}
		tr((s\mathbf{A} + (1 - s) \mathbf{JAJ})^{-1}) &= tr(\mathbf{J}((1 - s)\mathbf{A} + s \mathbf{JAJ})^{-1}\mathbf{J})\\
		& = tr(((1 - s)\mathbf{A} + s \mathbf{JAJ})^{-1}). 
		\end{aligned}
		\end{equation}
		
		According to the convexity, the minimum point for $f(s)$ is unique and cannot be any point other than $s = 1/2$ because of the symmetry, i.e.,
		\begin{equation}
		\begin{aligned}
		\mathrm{MSE} = \frac{1}{K\gamma}tr(\mathbf{A}^{-1}) \geqslant \frac{1}{K\gamma}tr\left(\left(\frac{\mathbf{A} + \mathbf{JAJ}}{2}\right)^{-1}\right).
		\end{aligned}
		\label{sym_eq}
		\end{equation}
		
		If $\mathbf{A} \neq \mathbf{JAJ}$, $tr(\mathbf{A}^{-1}) > tr\left(\left(\frac{\mathbf{A + JAJ}}{2}\right)^{-1}\right)$ according to the strict convexity of $f(s)$. In other words, $tr(\mathbf{A}^{-1}) = tr\left(\left(\frac{\mathbf{A + JAJ}}{2}\right)^{-1}\right)$ if and only if $\mathbf{A} = \mathbf{JAJ}$, which means that MSE will be minimized if and only if $\delta_i = \delta_{K - i}$, $i = 1, 2, \dots, K - 1$. The proof is complete.
\end{IEEEproof}

Based on the above two lemmas, we will prove Theorem \ref{equal_divided_within_group} for the most complicated case, $K = 7$. The other cases can be proved using the same method.


For $K = 7$, there are 6 variables. According to Lemma \ref{symmetricEq}, the number of variables is reduced by half since $\delta_1 = \delta_6$, $\delta_2 = \delta_5$, and $\delta_3 = \delta_4$. Then, $\mathbf{A}$ can be written as


\begin{figure*}[t b]
	\centering
	\begin{align}\label{AK=7}
	&\mathbf{A} = \left[ \begin{smallarray}{ccc;{2pt/2pt}c;{2pt/2pt}ccc}\small
	1 &1\!-\!\delta_1 &1\!-\!\delta_1\!-\!\delta_2 &1\!-\!\delta_1\!-\!\delta_2\!-\!\delta_3\! &1\!-\!\delta_1\!-\!\delta_2\!-\!2\delta_3 &1\!-\!\delta_1\!-\!2\delta_2\!-\!2\delta_3\! &1\!-\!2\delta_1\!-\!2\delta_2\!-\!2\delta_3\\
	1\!-\!\delta_1\! &1\! &1\!-\!\delta_2\! &1\!-\!\delta_2\!-\!\delta_3\! &1\!-\!\delta_2\!-\!2\delta_3\! &1\!-\!2\delta_2-\!2\delta_3\! &1\!-\!\delta_1\!-\!2\delta_2\!-\!2\delta_3\\
	1\!-\!\delta_1\!-\!\delta_2\! &1\!-\!\delta_2\! &1\! &1\!-\!\delta_3\! &1\!-\!2\delta_3\! &1\!-\!\delta_2\!-\!2\delta_3\! &1\!-\!\delta_1\!-\!\delta_2\!-\!2\delta_3\\ \hdashline[2pt/2pt]
	1\!-\!\delta_1\!-\!\delta_2\!-\!\delta_3\! &1\!-\!\delta_2\!-\!\delta_3\! &1\!-\!\delta_3\! &1\! &1\!-\!\delta_3\! &1\!-\!\delta_2\!-\!\delta_3\! &1\!-\!\delta_1\!-\!\delta_2\!-\!\delta_3\\ \hdashline[2pt/2pt]
	1\!-\!\delta_1\!-\!\delta_2\!-\!2\delta_3\! &1\!-\!\delta_2\!-\!2\delta_3\! &1\!-\!2\delta_3\! &1\!-\!\delta_3\! &1\! &1\!-\!\delta_2\! &1\!-\!\delta_1\!-\!\delta_2\\
	1\!-\!\delta_1\!-\!2\delta_2\!-\!2\delta_3\! &1\!-\!2\delta_2\!-\!2\delta_3\! &1\!-\!\delta_2\!-\!2\delta_3\! &1\!-\!\delta_2\!-\!\delta_3\! &1\!-\!\delta_2\! &1\! &1\!-\!\delta_1\\
	1\!-\!2\delta_1\!-\!2\delta_2\!-2\delta_3\! &1\!-\!\delta_1\!-\!2\delta_2\!-\!2\delta_3\! &1\!-\!\delta_1\!-\!\delta_2\!-\!2\delta_3\! &1\!-\!\delta_1\!-\!\delta_2\!-\!\delta_3\! &1\!-\!\delta_1\!-\!\delta_2\! &1\!-\!\delta_1\! &1\\
	\end{smallarray}\right]\notag\\
	= & \begin{pmat}[{||.}]
	\mathbf{B}\ &\mathbf{x}\ &\mathbf{C}^T\cr\-
	\mathbf{x}^T\ &q\ &\mathbf{x}^T\mathbf{J}\cr\-
	\mathbf{C}\ &\mathbf{Jx}\ &\mathbf{JBJ}\cr
	\end{pmat}.
	\end{align}

			\hrulefill
\end{figure*}

Note that the matrix in Eq. (\ref{AK=7}) is a symmetric centrosymmetric matrix which satisfies $\mathbf{JAJ = A}$. According to \cite{cantoni1976eigenvalues}, if the dimension is odd, any symmetric centrosymmetric matrix can be written as a partitioned matrix written in Eq. (\ref{AK=7}). There is a similar partitioning expression if the dimension is even, which can be used if $K$ is even. 

According to Theorem 2 in \cite{cantoni1976eigenvalues}, the eigenvalues of $\mathbf{A}$ are a combination of the eigenvalues of $\mathbf{A}_1 = \mathbf{B - JC}$ and the eigenvalues of the following matrix:
\begin{equation}
\begin{aligned}
\mathbf{A}_2 = \begin{bmatrix}
\mathbf{B + JC}\ &\sqrt{2}\mathbf{x}\\
\sqrt{2}\mathbf{x}^T\ &q
\end{bmatrix}.
\end{aligned}
\end{equation}

Assume that the eigenvalues of $\mathbf{A}_1$ are $\lambda_1, \lambda_2, \lambda_3$ and the eigenvalues of $\mathbf{A}_2$ are $\lambda_4, \lambda_5, \lambda_6, \lambda_7$. Then,

\begin{equation}
\begin{aligned}
tr(\mathbf{A}^{-1}) &= \sum_{i = 1}^{7} \frac{1}{\lambda_i} = \frac{\lambda_1\lambda_2 + \lambda_2\lambda_3 + \lambda_1\lambda_3}{\lambda_1\lambda_2\lambda_3}\\ &+ \frac{\lambda_4\lambda_5\lambda_6 + \lambda_4\lambda_5\lambda_7 + \lambda_4\lambda_6\lambda_7 + \lambda_5\lambda_6\lambda_7}{\lambda_4\lambda_5\lambda_6\lambda_7}
\end{aligned}
\end{equation}

Note that $\lambda_1\lambda_2\lambda_3 = \mathrm{det}(\mathbf{A}_1)$ and $\lambda_4\lambda_5\lambda_6\lambda_7 = \mathrm{det}(\mathbf{A}_2)$. Also, one can calculate

\begin{equation}
\begin{aligned}
\lambda_1\lambda_2 + \lambda_1\lambda_3 + \lambda_2\lambda_3 &= \mathrm{det}(\mathbf{A}_1^{11}) + \mathrm{det}(\mathbf{A}_1^{22}) + \mathrm{det}(\mathbf{A}_1^{33})\\
&= 4(2\delta_2\delta_3 + \delta_1\delta_2 + 2\delta_1\delta_3),
\end{aligned}
\end{equation}
where $\mathbf{A}_1^{ii}$ is the $2\times 2$ submatrix obtained by deleting the $i$th row and $i$th column.

In addition, we can compute
\begin{equation}
\begin{aligned}
\lambda_1\lambda_2\lambda_3 = 8\delta_1\delta_2\delta_3.
\end{aligned}
\end{equation}

Similarly, $\lambda_4\lambda_5\lambda_6 + \lambda_4\lambda_5\lambda_7 + \lambda_4\lambda_6\lambda_7 + \lambda_5\lambda_6\lambda_7$ is equal to the coefficient of $\lambda$ with the power 1 in the characteristic polynomial times -1 because
\begin{equation}
\begin{aligned}
\left. \frac{\mathrm{d}f(\lambda)}{\mathrm{d}\lambda}\right\vert_{\lambda = 0} &= \left. \frac{\mathrm{d} (\lambda - \lambda_4)(\lambda - \lambda_5)(\lambda - \lambda_6)(\lambda - \lambda_7)}{\mathrm{d}\lambda}\right\vert_{\lambda = 0}\\
 &=  -(\lambda_4\lambda_5\lambda_6 + \lambda_4\lambda_5\lambda_7 + \lambda_4\lambda_6\lambda_7 + \lambda_5\lambda_6\lambda_7).
\end{aligned}
\label{65}
\end{equation}

We can calculate (\ref{65}) using
\begin{equation}
\begin{aligned}
&\lambda_4\lambda_5\lambda_6 + \lambda_4\lambda_5\lambda_7 + \lambda_4\lambda_6\lambda_7 + \lambda_5\lambda_6\lambda_7 \\
&= 3\delta_1^2\delta_2 + 3\delta_1\delta_2^2 - 3\delta_1\delta_2 +  2\delta_1^2\delta_3 + 2\delta_1\delta_3^2 - 2\delta_1\delta_3 \\
&+ 2\delta_2^2\delta_3 + 2\delta_2\delta_3^2 - 2\delta_2\delta_3 + 6 \delta_1\delta_2\delta_3,
\end{aligned}
\end{equation}

and
\begin{equation}
\begin{aligned}
\lambda_1\lambda_2\lambda_3\lambda_4 = 2\delta_1\delta_2\delta_3(\delta_1 + \delta_2 + \delta_3 - 1).
\end{aligned}
\end{equation}

Then,
\begin{equation}
\begin{aligned}
\mathrm{MSE} &= \frac{1}{7\gamma} tr(\mathbf{A}^{-1}) = \frac{1}{7\gamma}\sum_{i = 1}^{7}\frac{1}{\lambda_i}\\
&= \frac{1}{7\gamma}\left(\frac{2}{\delta_1} + \frac{2}{\delta_2} + \frac{2}{\delta_3} + \frac{1}{2(1 - \delta_1 - \delta_2 - \delta_3)}\right). 
\end{aligned}
\end{equation}

If we assume that $\sum_{n = 1}^{6}\delta_n = 2 \sum_{n = 1}^{3} \delta_n = \Delta < \Delta_i / T$, MSE can be rewritten as
\begin{equation}
\begin{aligned}
\mathrm{MSE} = \frac{1}{7\gamma} \left(\frac{2}{\delta_1} + \frac{2}{\delta_2} + \frac{2}{\Delta/2 - \delta_1 - \delta_2} + \frac{1}{2(1 - \Delta/2)}\right).
\end{aligned}
\end{equation}

In order to minimize MSE, the partial derivative of MSE with respect to $\delta_1, \delta_2$ should be set to be 0, i.e.,
\begin{equation}
\begin{aligned}
\frac{\partial \mathrm{MSE}}{\partial \delta_1} &= \frac{1}{7\gamma} \left(-\frac{2}{\delta_1^2} + \frac{2}{(\Delta/2 - \delta_1 - \delta_2)^2}\right) = 0
\end{aligned}
\label{partialderivative1}
\end{equation}
\begin{equation}
\begin{aligned}
\frac{\partial \mathrm{MSE}}{\partial \delta_2} &= \frac{1}{7\gamma} \left(-\frac{2}{\delta_2^2} + \frac{2}{(\Delta/2 - \delta_1 - \delta_2)^2}\right) = 0.
\end{aligned}
\label{partialderivative2}
\end{equation}

Solving Eqs. (\ref{partialderivative1}) and (\ref{partialderivative2}) results in $\delta_1 = \delta_2 = \delta_3 = \Delta / 6$. Then, the expression of MSE becomes
\begin{equation}
\begin{aligned}
\mathrm{MSE} = \frac{1}{7\gamma}\left(\frac{36}{\Delta} + \frac{1}{2 - \Delta}\right),
\end{aligned}
\end{equation}
\begin{equation}
\begin{aligned}
\frac{\mathrm{d\ MSE}}{\mathrm{d}\ \Delta} = \frac{1}{7\gamma} \left(-\frac{36}{\Delta^2} + \frac{1}{(2 - \Delta)^2}\right) < 0,\ \mathrm{for}\ \Delta \in (0, 1). 
\end{aligned}
\label{deriveMSEofDelta}
\end{equation}

Since MSE is a decreasing function of $\Delta$, to minimize MSE, we should have $\Delta = \Delta_i/T$. 
which means $\sum_{n = 1}^{K - 1}\delta_n$ is $\Delta_i/T$. Then, $\delta_1 = \delta_2 = \delta_3 = \Delta_i / 6T$.

Our proof for the case $K = 7$ is complete. It is simple to extend this proof to the other cases $2 \leqslant K < 7$. The difference lies in the dimension of $\mathbf{A}$. The calculation of eigenvalues will be much easier if the dimension of $\mathbf{A}$ is smaller. Therefore, we omit the proof for the other cases.


%
%

\ifCLASSOPTIONcaptionsoff
  \newpage
\fi



%
\bibliographystyle{ieeetr}
\bibliography{IEEEabrv,IEEEexample}

\end{document}